%% file: pFHTSSCL.tex
\documentclass[journal]{IEEEtran}

\IEEEsettopmargin{t}{0.75in}

\usepackage{makecell}
\usepackage{stfloats}
\usepackage{lipsum}
\usepackage{cite}
\usepackage{graphicx}
\usepackage{subcaption}
\usepackage{xcolor}
\usepackage{pgfplots}
\usepackage{multirow}
\usepackage{upgreek}
\usepackage{amssymb}
\usepackage{amsmath, mathtools}
\usepackage{cases}
\usepackage{pifont}
\usepackage{bm}
\usepackage{bbm}
\usepackage{amsthm}
\usepackage{tabularx}
\usepackage{booktabs}
\usepackage{array}
\usepackage{float}
\usepackage{xcolor}
\usepackage{algorithmic}
\usepackage[linesnumbered,ruled,vlined]{algorithm2e}
\usepackage{changes}
\usepackage{bbm}
\usepackage{soul}
\usepackage{physics}
\usepackage{tikz}
\usepackage{etoolbox}
\usepackage{scalerel}
\usepackage[section]{placeins}
\usepackage{url}

\usetikzlibrary{decorations,decorations.markings}
\usetikzlibrary{optics}
\usetikzlibrary{arrows,positioning,shapes.geometric,spy}

\newcolumntype{Y}{>{\centering\arraybackslash}X}

\newcommand{\fixme}[2]{\ifx&#2&{\leavevmode\color{red}#1}\else{\leavevmode\color{red}FIXME\{}#1{\leavevmode\color{red}\}}\footnote{{\leavevmode\color{red}#2}}\PackageWarning{Fixme}{#1: #2}\fi}

\SetCommentSty{mycommfont}

\DeclareMathOperator*{\argmin}{arg\,min}

\DeclareMathOperator*{\sgn}{sgn}

\allowdisplaybreaks

\begin{document}
	
	\title{Successive-Cancellation Decoding of {Reed-Muller} Codes with Fast Hadamard Transform}
	
	\author{
		Nghia~Doan, 
		Seyyed~Ali~Hashemi, 
		and Warren~J.~Gross 
	}
	
	\maketitle
	\begin{abstract}
		\textcolor{black}{A novel permuted fast successive-cancellation list decoding algorithm with fast Hadamard transform (FHT-FSCL) is presented. The proposed decoder initializes $L$ $(L\ge1)$ active decoding paths with $L$ random codeword permutations sampled from the full symmetry group of the codes. The path extension in the permutation domain is carried out until the first constituent RM code of order $1$ is visited. Conventional path extension of the successive-cancellation list decoder is then utilized in the information bit domain. The simulation results show that for a RM code of length $512$ with $46$ information bits, by running $20$ parallel permuted FHT-FSCL decoders with $L=4$, we reduce $72\%$ of the computational complexity, $22\%$ of the decoding latency, and $84\%$ of the memory consumption of the state-of-the-art simplified successive-cancellation decoder that uses $512$ permutations sampled from the full symmetry group of the code, with similar error-correction performance at the target frame error rate of $10^{-4}$.}
	\end{abstract}
	
	\IEEEpeerreviewmaketitle
	\section{Introduction} \label{sec:intro}
	
	
	{Reed-Muller} (RM) codes are a class of linear error-correction codes introduced by Muller \cite{Muller} and Reed \cite{Reed}. Under the factor-graph representation, RM codes are similar to polar codes \cite{arikan}, which are used in the fifth generation (5G) cellular communication standard. The main difference between RM and polar codes is that RM codes are constructed to maximize the minimum distance of all the codewords \cite{Muller, Reed}, while polar codes are constructed to minimize the error probability under successive-cancellation (SC) decoding \cite{PolarConst_Vardy, PolarConst_Trifonov} or SC-list (SCL) decoding \cite{Huang19, Liao20}. Therefore, under maximum likelihood (ML) decoding, RM codes achieve a better error-correction performance than polar codes. However, ML decoding is generally impractical due to its high computational complexity.
	
	RM codes can be decoded using a wide range of practical decoding algorithms as introduced in \cite{Reed, sidel1992decoding, Sakkour, dumer2004recursive, Dumer06}. Recently, a recursive projection-aggregation (RPA) decoding algorithm \cite{Ye20} has been proposed that outperforms the error-correction performance of the decoders in \cite{Reed, sidel1992decoding, Sakkour, dumer2004recursive, Dumer06}. RPA decoding relies on the code projections to recursively reduce the code order, where the fast Hadamard transform (FHT) algorithm is used to optimally decode the first-order RM codes \cite{FHTD}. To obtain the estimated codeword, RPA decoding aggregates the decoding outputs of various lower-order code projections based on a majority voting technique. It was shown in \cite{Ye20} that RPA decoding can obtain near ML decoding performance for short and low-order RM codes, which also outperforms the error probability of the polar-cyclic redundancy check (CRC) concatenated codes under SCL algorithm at various code lengths and code rates. However, the main problems associated with RPA decoding are the high computational complexity, which significantly increases with the increase of the code rate, and the recursive nature of the algorithm, which hinders an efficient hardware implementation of the RPA decoder \cite{Nazari}. 
	
	By sharing the same factor-graph representation with polar codes, RM codes can be decoded using the fast and low complexity decoding algorithms of polar codes, namely fast SC (FSC) and fast SCL (FSCL) decoding algorithms \cite{gabi_fast_pcd, Ali_FSSCL, Ardakani_TCOM}. \textcolor{black}{In \cite{Ghaddar20}, FHT is integrated into FSC (FHT-FSC) and FSCL (FHT-FSCL) decoding to improve the error-correction performance of FSC and FSCL decoders for polar and RM codes}. However, the error probability of FSC-based and FSCL-based decoding with a small list size is inferior to that of RPA decoding, rendering the FSC-based algorithms to be unsuitable for applications with stringent frame error rate (FER) requirements. Although the FER of the FSCL decoder can be improved by increasing the list size, to obtain an FER performance close to that of RPA decoding, the list size required by FSCL decoding is impractical for RM codes of lengths greater than $128$ \cite{fathollahi2020sparse}. 
	
	To improve the error probability of RM codes under SC-based decoding, the received channel output can be decoded using the permuted factor-graph representations of the code \cite{hussami2009performance, BPList, Mikhail19, CABPList, Doan_GLOBECOM_18, Doan_GLOBECOM_20}. In \cite{Ali_SP}, instead of performing the decoding on a list of factor-graph permutations, the authors provided a decoding algorithm that carefully selects a good factor-graph permutation on the fly, significantly improving the FER of SCL decoding with small list size. However, the FER of the decoder proposed in \cite{Ali_SP} is also inferior to that of RPA decoding. \textcolor{black}{It was observed in \cite{geiselhart2020automorphism} that utilizing the codeword permutations sampled from the full symmetry group of RM codes provides significant error-correction performance gains when compared to the permutations sampled from the factor-graph permutation group of the codes.
	The recursive list decoding (RLD) algorithm with factor-graph permutations (RLDP) introduced in \cite{Dumer06} performs permutation decoding until the first information bit is visited. Then, only the decoding operations in the information bit domain are carried out to select the best decoding paths. The RLDP decoder significantly improves the error-correction performance of RLD decoding while relatively maintaining a similar computational complexity \cite{Dumer06}. It was shown in \cite{geiselhart2020automorphism} that the permuted SC-based decoders in \cite{geiselhart2020automorphism} can obtain similar or better error-correction performance when compared with RPA and RLDP decoders, while requiring significantly smaller computational complexity and decoding latency.}
	
	\textcolor{black}{In this paper, we propose efficient decoding techniques for FSCL decoding with FHT that provide better error-correction performance and complexity trade-offs when compared to the state-of-the-art RM decoder introduced in \cite{ geiselhart2020automorphism} and the RPA-based decoders in \cite{Ye20, fathollahi2020sparse}. Our contributions are summarized as follows.}
	\begin{enumerate}
		\item \textcolor{black}{We propose a novel permuted FHT-FSCL (p-FHT-FSCL) decoding algorithm of RM codes to significantly improve the error-correction performance of the FHT-FSCL decoder \cite{Ghaddar20}. In particular, the proposed p-FHT-FSCL decoder first initializes $L$ $(L\ge1)$ decoding paths with $L$ random codeword permutations sampled from the full symmetry group of RM codes. The path extension is carried out in the permutation domain to select the $L$ best decoding paths until the first constituent RM code of order $1$ is visited. Then, the conventional path extension in the information bit domain is considered to select the $L$ best decoding paths, while maintaining the previously selected codeword permutations of all the active paths. In addition, the proposed p-FHT-FSCL decoder utilizes an efficient path metric computation scheme to significantly reduce the computational complexity of FHT list decoding.}
		
		\item \textcolor{black}{We utilize the rich symmetry group of the codes to further improve the error-correction performance of p-FHT-FSCL decoding. Specifically, since each p-FHT-FSCL decoder utilizes random subsets of codeword permutations sampled from the full symmetry group of the codes, we run $M$ $(M>1)$ p-FHT-FSCL decoders with list size $L$ in parallel (p-FHT-FSCL-$L$-$M$) and select the most likely codeword from the set of $M$ candidate codewords.}
	\end{enumerate}
	\textcolor{black}{We numerically demonstrate that for various RM code configurations, the proposed decoders can obtain similar or better error-correction performance compared to that of the state-of-the-art permuted successive-cancellation decoder \cite{geiselhart2020automorphism}, the FHT-FSCL decoder \cite{Ghaddar20}, and RPA-based decoders \cite{Ye20, fathollahi2020sparse}, while significantly reducing the computational complexity, decoding latency, and memory requirements.}
	
	The remainder of this paper is organized as follows. Section~\ref{sec:pre} introduces background on RM codes and their decoding algorithms. Section~\ref{sec:prop} provides the details of the proposed decoding techniques, followed by a detailed numerical analysis. Finally, concluding remarks are presented in Section~\ref{sec:conclud}.
	
	\section{Preliminaries}
	\label{sec:pre}
	
	Throughout this paper, boldface letters indicate vectors and matrices. Unless otherwise specified, non-boldface letters indicate either binary, integer or real numbers. Greek letters are used to denote a RM code (node), the log-likelihood ratio (LLR) values, and the hard decisions associated with a RM code. Finally, sets are denoted by blackboard bold letters, e.g., $\mathbb{R}$ is the set containing real numbers.
	
	\subsection{{Reed-Muller} Codes}
	A RM code is specified by a pair of integers $r$ and $m$, ${0 \leq r \leq m}$, and is denoted as $\mathcal{RM}(r,m)$, where $r$ is the order of the code. $\mathcal{RM}(r,m)$ has a code length ${N=2^m}$ with ${K=\sum_{i=0}^{r} {m \choose i}}$ information bits, and a minimum distance ${d=2^{m-r}}$. Note that $\mathcal{RM}(m,m)$ is a rate-1 code that contains all the $2^N$ binary codewords of length $N$, and $\mathcal{RM}(-1,m)$ is a rate-0 code that contains the all-zero codeword of size $N$. A RM code can be constructed by applying a linear transformation 
	to the binary message word $\bm{u} = \{u_0,u_1,\ldots,u_{N-1}\}$ as $\bm{x} = \bm{u}\bm{G}^{\otimes m}$ where $\bm{x} = \{x_0,x_1,\ldots,x_{N-1}\}$ is the codeword and $\bm{G}^{\otimes m}$ is the $m$-th Kronecker power of the matrix $\bm{G}=\bigl[\begin{smallmatrix} 1&0\\ 1&1 \end{smallmatrix} \bigr]$ \cite{Arikan10}. The element $u_i$ of $\bm{u}$ is fixed to $0$ if the weight of the $i$-th row of $\bm{G}^{\otimes m}$, denoted as $w_i$, is smaller than $d$. Formally, $u_i = 0$ $\forall i \in \mathbbm{F}$, where $\mathbbm{F} = \{i|0\leq i < N, w_i < d\}$. In addition, we denote by $\mathbbm{I}$ the set of information bits, i.e., $\mathbbm{I} = \{i|0\leq i < N, w_i \ge d\}$, and the sets $\mathbbm{I}$ and $\mathbbm{F}$ are known to both the encoder and the decoder. 
	
\begin{figure}[t]
		\centering
		\begin{subfigure}{1\columnwidth}
			\centering
			\input{Figures/PolarFactorGraph.tikz}
			\caption{}
		\end{subfigure}
		
		\begin{subfigure}{1\columnwidth}
			\centering
			\hspace*{20pt} \input{Figures/SCPE.tikz}
			\caption{}
		\end{subfigure}
		
		\caption{(a) Factor-graph representation of $\mathcal{RM}(1,3)$, and (b) a processing element.}
		\label{fig:rm:fg}
		\vspace*{-0.5\baselineskip}
	\end{figure}
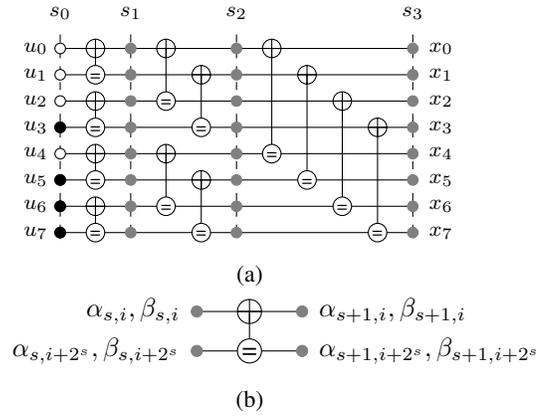

	In this paper, the codeword $\bm{x}$ is modulated using binary phase-shift keying (BPSK) modulation, and additive white Gaussian noise (AWGN) channel model is considered. Therefore, the soft vector of the transmitted codeword received by the decoder is given as ${\bm{y}=(\mathbf{1}-2\bm{x})+\bm{z}}$, where $\mathbf{1}$ is an all-one vector of size $N$, and $\bm{z} \in \mathbbm{R}^N$ is a Gaussian noise vector with variance $\sigma^2$ and zero mean. In the log-likelihood ratio (LLR) domain, the LLR vector of the transmitted codeword is given as
	${\bm{\alpha}_m=\ln{\frac{Pr(\bm{x}=0|\bm{y})}{Pr(\bm{x}=1|\bm{y})}}=\frac{2\bm{y}}{\sigma^2}}$. Fig.~\ref{fig:rm:fg}(a) illustrates the encoding process of $\mathcal{RM}(1,3)$ using the factor-graph representation of the code, where $N=8$, $K=4$, and $\mathbbm{I}=\{3,5,6,7\}$ \cite{Arikan08}.
	
	\subsection{Successive-Cancellation and Successive-Cancellation List Decoding} 
	SC decoding is executed on the factor-graph representation of the code \cite{Dumer06, arikan}. To obtain the message word, the soft LLR values and the hard bit estimations are propagated through all the processing elements (PEs), which are depicted in Fig.~\ref{fig:rm:fg}(b). Each PE performs the following computations: $\alpha_{s,i} = f(\alpha_{s+1,i},\alpha_{s+1,i+2^s})$ and $\alpha_{s,i+2^s} = g(\alpha_{s+1,i},\alpha_{s+1,i+2^s},\beta_{s,i})$, where $\alpha_{s,i}$ and $\beta_{s,i}$ are the soft LLR value and the hard-bit estimation at the $s$-th stage and the $i$-th bit, respectively. The min-sum approximation formulations of $f$ and $g$ are $f(a,b) = \min(|a|,|b|)\sgn(a)\sgn(b)$, and $g(a,b,c) = b + (1-2c)a$. The soft LLR values at the $m$-th stage are initialized to $\bm{\alpha}_m$ and the hard-bit estimation of an information bit at the $0$-th stage is obtained as $\hat{u}_{i} = \beta_{0,i}=\frac{1 - \sgn(\alpha_{0,i})}{2}$, $\forall i \in \mathbbm{I}$. The hard-bit values of the PE are then computed as $\beta_{s+1,i} = \beta_{s,i} \oplus \beta_{s,i+2^s}$ and $\beta_{s+1,i+2^s} = \beta_{s,i+2^s}$.
	
	
	Although SC decoding is a low-complexity decoding algorithm, its FER performance for short to moderate code lengths is mediocre. Therefore, SCL decoding was introduced to significantly improve the error-correction performance of SC decoding \cite{Dumer06, tal_list, Alexios_LLR_SCLD}. Under SCL decoding, the estimation of an information bit $\hat{u}_i$ $(i \in \mathbbm{I})$ is considered to be both $0$ and $1$, causing a path splitting and doubling the number of candidate codewords (decoding paths) after each split. To prevent the exponential growth of the number of decoding paths, a path metric is utilized to select the $L$ most probable decoding paths after each information bit is decoded. In the LLR domain, the low-complexity path metric is obtained as \cite{Alexios_LLR_SCLD}
	\begin{equation}
		\label{equ:polar:PM_LLR}
		\textup{PM}_l=
		\begin{cases}
			\textup{PM}_l + \abs{\alpha_{{0,i}_l}} & \text{ if } \hat{u}_i \neq \frac{1-\sgn(\alpha_{{0,i}_l})}{2},\\
			\textup{PM}_l & \text{ otherwise,}
		\end{cases}
	\end{equation}
	where $\alpha_{{0,i}_l}$ denotes the soft value of the $i$-th bit at stage $0$ of the $l$-th path. Initially, $\textup{PM}_l=0$, $\forall l$. After each information bit is decoded, only $L$ paths with the smallest path metric values are kept to continue the decoding. At the end of the decoding process, only the decoding path that has the smallest path metric is selected as the decoding output. 
	
	\subsection{\textcolor{black}{Fast Successive-Cancellation List Decoding of {Reed-Muller} Codes}} 
	\label{sec:polar:FSCL}
		
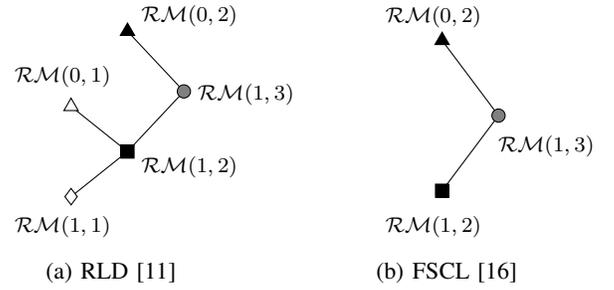
\begin{figure}
		\begin{subfigure}{0.49\linewidth}
			\centering
			\input{Figures/RLD.tikz.tex}
			\caption{RLD \cite{Dumer06}}
		\end{subfigure}
		\begin{subfigure}{0.49\linewidth}
			\centering
			\input{Figures/FSCL.tikz.tex}
			\caption{FSCL \cite{Ali_FSSCL}}
		\end{subfigure}
		\caption{Binary tree representations of $\mathcal{RM}(1,3)$ under (a) RLD \cite{Dumer06} and (b) FSCL \cite{Ali_FSSCL}.}
		\label{fig:rm:tree}
	\end{figure}

	SCL decoding can also be illustrated using a binary tree representation \cite{dumer2004recursive, Dumer06, Ali_FSSCL, Ardakani_TCOM}. Fig.~\ref{fig:rm:tree}(a) shows a full binary tree representation of $\mathcal{RM}(1,3)$, whose factor graph is depicted in Fig.~\ref{fig:rm:fg}(a). In \cite{Ali_FSSCL, Ardakani_TCOM}, the authors proposed the FSCL decoding algorithms for various polar subcodes that can be directly applied to RM codes. For some special nodes, FSCL decoding preserves the error-correction performance of SCL decoding while completely removing the need to visit the descendant nodes. Therefore, the decoding latency of the FSCL algorithm is significantly reduced when compared with SCL decoding.
	
	Consider a parent node $\nu$ located at the $s$-th stage $(s>0)$ of the binary tree, which is a $\mathcal{RM}(r_\nu, m_\nu)$. There are $N_\nu$ LLR values and $N_\nu$ hard decisions associated with this node, where $N_\nu=2^{m_\nu}=2^s$. Let $\bm{\alpha}_{\nu_l}$ and $\bm{\beta}_{\nu_l}$ be the soft and hard values associated with a parent node $\nu$ of the $l$-th decoding path, respectively. $\bm{\alpha}_{\nu_l}$ and $\bm{\beta}_{\nu_l}$ are given as
	\begin{equation*}
		\begin{cases}
			\bm{\alpha}_{\nu_l} = \{\alpha_{{s,i_{{\min}_{\nu_l}}}},\ldots,\alpha_{{s,i_{{\max}_{\nu_l}}}}\},\\
			\bm{\beta}_{\nu_l} = \{\beta_{{s,i_{{\min}_{\nu_l}}}},\ldots,\beta_{{s,i_{{\max}_{\nu_l}}}}\},\\
		\end{cases}
	\end{equation*}
	where $i_{{\min}_{\nu_l}}$ and $i_{{\max}_{\nu_l}}$ are the bit indices such that $0 \leq i_{{\min}_{\nu_l}} < i_{{\max}_{\nu_l}} \leq N-1$ and $i_{{\max}_{\nu_l}} - i_{{\min}_{\nu_l}} = N_\nu-1$. The hard-decision values of $\nu$ in the bipolar form are denoted as $\bm{\eta}_{\nu_l} = \{\eta_{{s,i_{{\min}_{\nu_l}}}},\ldots,\eta_{{s,i_{{\max}_{\nu_l}}}}\}$, where $\eta_{s,i}=1-2\beta_{s,i}$, $i_{{\min}_{\nu_l}} \leq i \leq i_{{\max}_{\nu_l}}$.
	
	Let $\tau$ be the minimum number of path splittings that allows FSCL decoding to preserve the error-correction performance of the conventional SCL decoding algorithm for the single-parity check (SPC) nodes. Note that all the leaf nodes of an SPC node are information bits, except for $\beta_{0,i_{{\min}_{\nu}}}$. Also, let the elements of $\bm{\alpha}_{\nu}$ corresponding to the SPC node be sorted in the following order: $\abs{\alpha_{{s,i_{{\min}_{\nu}}}}} \leq \abs{\alpha_{{s,i_{{\min}_{\nu}}+1}}} \leq \ldots \leq \abs{\alpha_{{s,i_{{\max}_{\nu}}}}}$. The decoding operations of $\mathcal{RM}(m_\nu-1, m_\nu)$ (SPC node) under FSCL decoding are summarized as follows.
	
	The parity check sum of the $l$-th path is first obtained as \cite{Ali_FSSCL, Ardakani_TCOM}
	\begin{equation}
		\label{equ:Parity}
		p_l=\bigoplus_{i=i_{\min_{\nu_l}}}^{i_{\max_{\nu_l}}}\frac{1-\sgn(\alpha_{s,i})}{2}.
	\end{equation}
	The path metric is then updated as \cite{Ali_FSSCL, Ardakani_TCOM}
	\begin{equation}
		\textup{PM}_l=
		\begin{cases}
			\textup{PM}_l + \abs{\alpha_{s,i_{\min_l}}} & \text{ if } p=1,\\
			\textup{PM}_l & \text{ otherwise.}\\
		\end{cases}
	\end{equation}
	The decoding continues with $\tau$ path splittings, where ${\tau=\min(L,N_\nu)}$. In each new path splitting, the path metric is updated as \cite{Ali_FSSCL, Ardakani_TCOM}
	\begin{equation}
		\textup{PM}_l=
		\begin{cases}
			\textup{PM}_l + \abs{\alpha_{{s,i}}} +(1-2p_{s,{i-1}})\abs{\alpha_{s,i_{\min_l}}}\\ 
			\hspace*{70pt} \text{if } \eta_{s,i} \neq \sgn(\alpha_{s,i}),\\
			\textup{PM}_l \hspace*{53.5pt} \text{otherwise,}\\
		\end{cases}
	\end{equation}
	where $i_{\min_l} < i \leq i_{\min_l}+\tau$. The parity check sum is then updated after each path splitting as \cite{Ardakani_TCOM}
	\begin{equation}
		\label{equ:FSSCL-SPC:gamma_update}
		p_l=
		\begin{cases}
			1\oplus p_l &\text{if } \eta_{s,i_l} \neq \sgn(\alpha_{s,i_l}),\\
			p_l &\text{otherwise.}\\
		\end{cases}
	\end{equation}
	When all the bits are estimated, the hard decision of the least reliable bit is updated to maintain the parity check condition of the SPC node, which is given as \cite{Ali_FSSCL, Ardakani_TCOM}
	\begin{equation}
		\label{equ:FSSCL-SPC:beta_update}
		\beta_{s,i_{\min_{\nu_l}}}= \bigoplus_{i=i_{\min_{\nu_l}}+1}^{i_{\max_{\nu_l}}}\beta_{s,i}.
	\end{equation}
	
	\textcolor{black}{The RLD algorithms proposed in \cite{dumer2004recursive, Dumer06} considers fast decoding for $\mathcal{RM}(0,m_\nu)$, $\mathcal{RM}(1,m_\nu)$, and $\mathcal{RM}(m_\nu,m_\nu)$ (rate-1). In \cite{Ghaddar20}, FHT is integrated into FSCL decoding (FHT-FSCL) to improve the error-correction performance of FSCL decoding for polar and RM codes. Fig.~\ref{fig:rm:tree} shows an example of the binary tree representations used by RLD \cite{Dumer06} and FSCL \cite{Ali_FSSCL} for decoding $\mathcal{RM}(1,3)$. Note that under FHT-FSCL decoding \cite{Ghaddar20}, $\mathcal{RM}(1,3)$ is directly decoded using FHT without decomposing the code into smaller RM codes.}
	
	The memory requirements of FHT-FSC and FHT-FSCL decoding algorithms are given as \cite{Ghaddar20, Ali_FSSCL}
	\begin{equation}
		\mathcal{M}_\text{FHT-FSC} = \underbrace{(2N-1)Q}_\text{$\bm{\alpha}$ memory}+\underbrace{N}_\text{$\bm{\beta}$ memory},
	\end{equation}
	and
	\begin{equation}
		\label{equ:mem:FSCL}
		\begin{split}
			\mathcal{M}_\text{FHT-FSCL} &= \underbrace{(N+(N-1)L)Q}_{\text{$\bm{\alpha}$ memory}}+\underbrace{LQ}_{\text{PM memory}}+\underbrace{2NL}_\text{$\bm{\beta}$ memory}\\
			&= N(L+1)Q+2NL,
		\end{split}
	\end{equation}
	where $Q$ is the number of bits that are used to quantize the LLR and path metric values. In addition, we quantify the computational complexity of the FSC-based and FSCL-based decoders by counting the number of floating-point additions and comparisons required by the LLR sorting of the SPC nodes and the path metric sorting of each path split.
	
	\subsection{Recursive Projection Aggregation Decoding}
	\label{sec:pre:RPA}
		
	RPA decoding is an algorithm that can achieve near-ML decoding performance for low-order RM codes of short to moderate code lengths \cite{Ye20}. The RPA decoding algorithm performs iterative decoding operations on a set $\mathbb{B}$ that contains $2^m-1$ one-dimensional code projections of $\mathcal{RM}(r,m)$, where the projected first-order RM codes are optimally decoded using the FHT algorithm \cite{FHTD}. Algorithm~\ref{algo:RPA} provides the details of the RPA decoder introduced in \cite{Ye20}. 
	
\begin{figure}
		\centering
		\input{Figures/projection.tikz}
		\caption{Example of the one-dimensional projection set $\mathbb{B}$ of a RM code of length $8$. An element of $\mathbb{B}$ is indexed as $\mathbb{B}_{j,k,z}$ where $0 \leq j \leq 6$, $0 \leq k \leq 3$, and $0 \leq z \leq 1$.}
		\label{fig:projections}
	\end{figure}
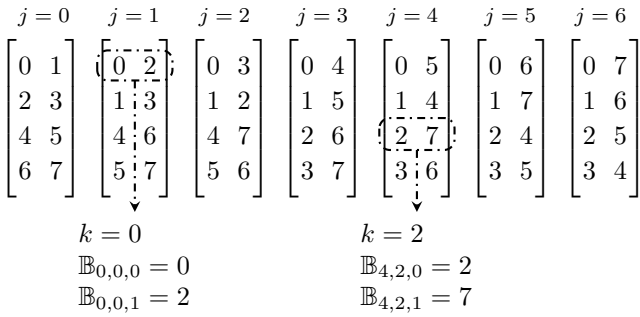

	Given $\mathcal{RM}(r,m)$ with $r>1$ and the channel LLR vector $\bm{y}$, the RPA algorithm performs a maximum number of $\big \lceil \frac{m}{2} \big \rceil$ iterations. At each iteration $i$ $(0 \leq i < \big \lceil \frac{m}{2} \big \rceil)$, the decoder projects the channel LLR vector $\bm{y}$ into the  LLR vector $\bm{y}_{\mathbb{B}_j}$ corresponding to a projected $\mathcal{RM}(r-1,m-1)$ code by using the $j$-th projection of $\mathbb{B}$ \cite{Ye20}. In this paper, we encode the one-dimensional projection set $\mathbb{B}$ as a matrix of the elements $\mathbb{B}_{j,k,z}$. Specifically, $\mathbb{B}_{j,k,z}$ is the $z$-th element of the $k$-th coset that belongs to the $j$-th projection of $\mathbb{B}$, where $0 \leq z < 2$, $0 \leq k \leq 2^{m-1}$, and $0 \leq j \leq 2^m-1$. Fig.~\ref{fig:projections} shows an example of the projection set $\mathbb{B}$ of a RM code of length $8$.
	
	In Algorithm~\ref{algo:RPA}, the projected LLR vector $\bm{y}_{\mathbb{B}_j}$ is first initialized as an all-zero vector of size $2^{m-1}$. The $k$-th element of $\bm{y}_{\mathbb{B}_j}$ is then calculated using the $\texttt{Projection}(\cdot)$ function (line 9, Algorithm~\ref{algo:RPA}), where $\texttt{Projection}(a,b)$ computes $\ln(\exp(a + b) + 1) - \ln(\exp(a) + \exp(b))$ \cite{Ye20}. The estimated hard decision values of $\bm{y}_{\mathbb{B}_j}$ are recursively calculated by running the $\texttt{RPA}(\cdot)$ decoding function for $\bm{y}_{\mathbb{B}_j}$, followed by the aggregation step that updates the aggregated LLR vector $\bm{y}_\text{agg}$ using the estimated values of $\hat{\bm{x}}_{\mathbb{B}_j}$ and the channel LLR vector $\bm{y}$ (line 12 and 13, Algorithm~\ref{algo:RPA}) \cite{Ye20}. To reduce the decoding latency of RPA decoding, a convergence condition of $\bm{y}$ and $\bm{y}_\text{agg}$ is verified at each iteration (line 14, Algorithm~\ref{algo:RPA}), where $\Delta$ is a scaling factor used to ensure that $\bm{y}_\text{agg}$ reaches a stable state \cite{Ye20}. If the termination condition is satisfied or the maximum number of iterations has been reached, the RPA decoder returns the estimated codeword by making the hard decisions from $\bm{y}_\text{agg}$, otherwise $\bm{y}$ is updated as $\bm{y}_\text{agg}$ and the decoder performs the next decoding iteration.
	
	\begin{algorithm}[t]
		\DontPrintSemicolon
		\caption{The $\texttt{RPA}(\cdot)$ Decoding Algorithm \cite{Ye20}}
		\label{algo:RPA}
		\SetKwInOut{Input}{Input}
		\SetKwInOut{Output}{Output}
		\SetKwInput{kwIn}{in}	
		
		\Input{$r,m,\bm{y}$}
		\Output{$\hat{\bm{x}}$}
		\vskip 0.25cm

		\SetKwFunction{FMain}{RPA}
		\SetKwProg{Fn}{Function}{:}{}
		\If{$r=1$}{
			$\hat{\bm{x}} \leftarrow \text{Decode $\bm{y}$ using FHT \cite{FHTD}}$\\
		}
		\Else
		{
			\For{$i \leftarrow 0$ \KwTo $\big\lceil \frac{m}{2} \big\rceil - 1$}{
				$\bm{y}_\text{agg} \leftarrow \bm{0}$\\
				\For{$j \leftarrow 0$ \KwTo $2^m-2$}{
					\tcc{Projection}
					$\bm{y}_{\mathbb{B}_j} \leftarrow \bm{0}$\\
					\For{$k \leftarrow 0$ \KwTo $2^{m-1}-1$}{
						$\bm{y}_{\mathbb{B}_j}[k] \leftarrow \texttt{Project}(\bm{y}[\mathbb{B}_{j,k,0}], \bm{y}[\mathbb{B}_{j,k,1}])$
					}
					\tcc{Recursively decode $\bm{y}_{\mathbb{B}_j}$}
					$\hat{\bm{x}}_{\mathbb{B}_j} \leftarrow \texttt{RPA}(r-1,m-1,\bm{y}_{\mathbb{B}_j})$\\
					\tcc{Aggregration}
					\For{$k \leftarrow 0$ \KwTo $2^{m-1}-1$}{
						$\bm{y}_\text{agg}[\mathbb{B}_{j,k,0}] \leftarrow \bm{y}_\text{agg}[\mathbb{B}_{j,k,0}]+(1-2\hat{\bm{x}}_{\mathbb{B}_j}[k])\bm{y}[\mathbb{B}_{j,k,1}]$\\
						$\bm{y}_\text{agg}[\mathbb{B}_{j,k,1}] \leftarrow \bm{y}_\text{agg}[\mathbb{B}_{j,k,1}]+(1-2\hat{\bm{x}}_{\mathbb{B}_j}[k])\bm{y}[\mathbb{B}_{j,k,0}]$\\
					}
				}
				\tcc{Update $\bm{y}$ and check for termination}
				$\bm{y}_\text{agg} \leftarrow \frac{\bm{y}_\text{agg}}{2^m-1}$\\
				\If{$\abs{\bm{y}_\text{agg}[j]-\bm{y}[j]} \leq \Delta\abs{\bm{y}[j]}$ $\forall j$, $0 \leq j < 2^m-1$ \textup{\textbf{or}} $i = \big\lceil \frac{m}{2} \big\rceil - 1$}
				{
					$\hat{\bm{x}} \leftarrow$ Obtain hard decisions from $\bm{y}_\text{agg}$\\
					\Return{$\hat{\bm{x}}$}
				}
				\Else
				{
					$\bm{y} \leftarrow \bm{y}_\text{agg}$
				}
			}
		}
		\Return{$\hat{\bm{x}}$}
	\end{algorithm}
	
	In addition, we count the number of floating-point additions/subtractions and floating-point comparisons to quantify the computational complexity of the RPA algorithm. In particular, the $\texttt{Projection}(a,b)$ function requires $4$ additions/subtractions as we consider the transcendental computations used in $\texttt{Projection}(a,b)$ are implemented using a look-up-table (LUT) without degrading the FER performance. Each aggregation operation used in lines 12 and 13 of Algorithm~\ref{algo:RPA} requires $1$ addition. Furthermore, the FHT decoding algorithm, when applied to a $\mathcal{RM}(1,m)$, requires $m2^m$ additions and $2^m$ comparisons for the selection of the most probable codeword \cite{FHTD}. In this paper, we set $\Delta = 0.0625$ as opposed to $0.05$ \cite{Ye20}. Therefore, a multiplication with $\Delta$ can be implemented by a low-cost shift operation. Consequently, the verification of the termination condition in line 15 of Algorithm~\ref{algo:RPA} requires $2^{m+1}$ operations. 
	
	\textcolor{black}{A variant of RPA decoding, referred as sparse RPA (SRPA) decoding, was proposed in \cite{fathollahi2020sparse} that reduces the decoding complexity of the RPA decoder. Specifically, the SRPA decoding algorithm runs two fully-parallel RPA decoders with each decoder using a quarter of the code projections at each recursion step \cite{fathollahi2020sparse}. Thus, the SRPA decoder effectively reduces $50\%$ of the total number of projections used by the conventional RPA decoder \cite{Ye20}. This configuration incurs negligible error-correction performance loss with respect to the conventional RPA decoder in \cite{Ye20} for the second and third order RM codes of size $256$. In this paper, we consider a fully-parallel implementation for the RPA-based decoders, in which all the operations that can be carried out concurrently are executed at the same time.}
	
	\section{\textcolor{black}{Permuted Fast SCL Decoding with Fast Hadamard Transform}}
	\label{sec:prop}
	
	\subsection{\textcolor{black}{Proposed Decoding Algorithm}}
	\label{sec:prop:main}
	\textcolor{black}{The FHT-FSCL decoder introduced in \cite{Ghaddar20} provides a better error-correction performance in comparison with the FSCL decoder for RM codes of low orders when small list sizes ($L\leq 8$) are used. However, for RM codes of orders greater than $2$ and with a relatively large list size $(L>8)$, FHT-FSCL decoding provides a negligible error-correction performance gain compared to FSCL decoding. Inspired by the previous RM decoders introduced in \cite{Dumer06,Ali_SP} and \cite{geiselhart2020automorphism}, in this section, we propose a permuted FHT-FSCL decoding algorithm that significantly improves the FER performance of FHT-FSCL decoding, while relatively maintaining the computational complexity, decoding latency, and memory requirement of FHT-FSCL decoding when the same list size is used. The details of the proposed p-FHT-FSCL decoder with list size $L\ge1$ are provided in Algorithm~\ref{algo:p-FHT-FSCL}.}
		
	\begin{algorithm}[t!]
		\DontPrintSemicolon
		\caption{$\texttt{p-FHT-FSCL}(\cdot)$ Decoding}
		\label{algo:p-FHT-FSCL}
		\SetKwInOut{Input}{Input}
		\SetKwInOut{Output}{Output}
		\SetKwInput{kwIn}{in}
		\SetKwFunction{pFHTFSCL}{p-FHT-FSCL}
		\SetKwFunction{FHTL}{FHTL}
		\SetKwFunction{SPCL}{SPCL}
		\SetKwProg{Fn}{Function}{:}{}
		\Input{$\{\bm{\alpha}_{\nu_l}, \textup{PM}_l\}_{0 \leq l < L}$}		
		\Output{$\{\hat{\bm{x}}_{\nu_l}, \textup{PM}_l\}_{0 \leq l < L}$}
		\Fn{$\pFHTFSCL\left(\{\bm{\alpha}_{\gamma_l}, \textup{PM}_l\}_{0 \leq l < L}\right)$}{
			\tcc{Initialize $L$ decoding paths with $L$ random codeword permutaitons once}
			\If{\textup{$\mathcal{RM}(r_\nu,m_\nu)$ is the root node}}{
				$\textup{isPermutation}\leftarrow\text{True}$\\
				$\{\pi^\textup{init}_{l}: \bm{\alpha}_{\nu_l} \xrightarrow{\pi^\textup{init}_{l}} \bm{\alpha}_{\nu_l}\}_{0 \leq l < L}$\\
			}
			\If{$1<r_\nu<m_\nu-1$}
			{
				\tcc{Permutation decoding if applicable}
				\If{$\textup{isPermutation}=\textup{True}$}{
					$\textup{isPer}_\nu=\textup{True}$\\
					\For{$l \leftarrow 0$ \KwTo $2L$}{
						$l_\text{org} \leftarrow l\mod L$;
						$\pi_{l}:$ $\bm{\alpha}_{\nu_{l_\text{org}}} \xrightarrow{\pi_{l}} \bm{\alpha}^{\textup{tmp}}_{\nu}$\\
						$\bm{\alpha}^\textup{tmp}_{\lambda} \leftarrow f\left(\bm{\alpha}^{\textup{tmp}}_{\nu}\right)$; $\textup{LM}_l\leftarrow\sum_{\forall i}\abs{{\alpha}^\textup{tmp}_{\lambda}[i]}$\\
					}
					$\{\pi^*_l, l^*_\textup{org}\}_{0 \leq l < L} \leftarrow \texttt{Sort}\left(\textup{LM}_0,\ldots,\textup{LM}_{2L-1}\right)$\\
					$\{\pi^*_l: \bm{\alpha}_{\nu_{l^*_\textup{org}}} \xrightarrow{\pi^*_l} \bm{\alpha}_{\nu_l}\}_{0 \leq l < L}$\\
				}
				\tcc{Recursively decode the left-child node $\mathcal{RM}(r_\nu-1,m_\nu-1)$}
				$\{\bm{\alpha}_{\lambda_l} \leftarrow f\left( \bm{\alpha}_{\nu_l} \right)\}_{0 \leq l < L}$\\
				$\{\hat{\bm{x}}_{\lambda_l}, \textup{PM}_l\}_{0 \leq l < L} \leftarrow \pFHTFSCL\left(\{\bm{\alpha}_{\lambda_l}, \textup{PM}_l\}_{0 \leq l < L}\right)$\\
				\tcc{Recursively decode the right-child node $\mathcal{RM}(r_\nu,m_\nu-1)$}
				$\{\bm{\alpha}_{\gamma_l} \leftarrow g( \bm{\alpha}_{\nu_{l_\textup{org}}},\hat{\bm{x}}_{\lambda_l})\}_{0 \leq l < L}$\\
				$\{\hat{\bm{x}}_{\gamma_l}, \textup{PM}_l\}_{0 \leq l < L} \leftarrow \pFHTFSCL\left(\{\bm{\alpha}_{\gamma_l}, \textup{PM}_l\}_{0 \leq l < L}\right)$\\
				\tcc{Form the estimation of $\mathcal{RM}(r_\nu,m_\nu)$ and repermute if applicable}
				$\{\hat{\bm{x}}_{\nu_l}\leftarrow\texttt{Concate}(\hat{\bm{x}}_{\gamma_l}\oplus\hat{\bm{x}}_{\lambda_{l_\textup{org}}},\hat{\bm{x}}_{\gamma_l})\}_{0 \leq l < L}$\\
				\If{$\textup{isPer}_\nu=\textup{True}$}{
					$\{(\pi^*_{l_\textup{org}})^{-1}: \hat{\bm{x}}_{\nu_l} \xrightarrow{(\pi^*_{l_\textup{org}})^{-1}} \hat{\bm{x}}_{\nu_l}\}_{0 \leq l < L}$
				}
				\If{\textup{$\mathcal{RM}(r_\nu,m_\nu)$ is the root node}}{
					$\{(\pi^\textup{init}_{l_\textup{org}})^{-1}: \hat{\bm{x}}_{\nu_l} \xrightarrow{(\pi^\textup{init}_{l_\textup{org}})^{-1}} \hat{\bm{x}}_{\nu_l}\}_{0 \leq l < L}$
				}
			}
			\ElseIf{$r=1$}{
				\If{$\textup{isPermutation}=\textup{True}$}{
					$\textup{isPermutation} \leftarrow \textup{False}$
				}
				$\{\hat{\bm{x}}_{\nu_l}, \textup{PM}_l\}_{0 \leq l < L} \leftarrow \FHTL\left(\{\bm{\alpha}_{\nu_l}, \textup{PM}_l\}_{0 \leq l < L}\right)$\\
			}
			\ElseIf{$r=m-1$}{
				$\{\bm{\alpha}_{\nu_l}, \textup{PM}_l\}_{0 \leq l < L} \leftarrow \SPCL\left(\{\hat{\bm{x}}_{\nu_l}, \textup{PM}_l\}_{0 \leq l < L}\right)$\\
			}
			\Return $\{\hat{\bm{x}}_{\nu_l}, \textup{PM}_l\}_{0 \leq l < L}$
		}
	\end{algorithm}

	\textcolor{black}{The proposed decoder is initialized with $L$ active decoding paths whose LLR vectors are set to the received channel LLRs $\bm{y}$, and the path metric is set to $0$. The proposed decoder then permutes the LLR vectors of the $L$ decoding paths using $L$ random codeword permutations sampled from the full symmetry group of the RM codes (lines 2-4 in Algorithm~\ref{algo:p-FHT-FSCL}), where $\pi^{\text{init}}_l$ indicates the initial codeword permutation applied to the $l$-th path \cite{geiselhart2020automorphism}. This initialization process is only performed once for each received channel LLR vector $\bm{y}$. The decoding continues with the path extension performed in the permutation domain until the first constituent RM code of order $1$ is visited. Lines 6-12 of Algorithm~\ref{algo:p-FHT-FSCL} specify the proposed permutation decoding. In particular, we sample two random codeword permutations, $\pi_l$, for each active decoding path to obtain the permutations of $\bm{\alpha}_{\nu_l}$, denoted as $\bm{\alpha}^\textup{tmp}_{\nu}$. The permuted LLR vector $\bm{\alpha}^\textup{tmp}_{\nu}$ is used to compute the LLR values of the left-child node $\bm{\alpha}^\textup{tmp}_{\lambda}$ using the $f(\cdot)$ function. Then, the reliability metric proposed in \cite{Ali_SP} is computed to select the $L$ permutations that have the maximum channel reliabilities $\textup{LM}_l$ of the left-child node $\lambda$ (lines 8-11 of Algorithm~\ref{algo:p-FHT-FSCL}). In line 12 of Algorithm~\ref{algo:p-FHT-FSCL}, the selected permutations are applied to the input LLR vectors to form the $L$ best decoding paths. Here, by $l^*_\textup{org}$ we denote the index of the input LLR vector whose permutation $\pi^*_l$ provides the channel reliability that is among the $L$ largest channel reliabilities.}
	
	\textcolor{black}{Note that the proposed permutation decoding selects the best permutations originated from all the current active decoding paths, which is different from the decoders proposed in \cite{Ali_SP, Dumer06, geiselhart2020automorphism} where permutation decoding is utilized separately for each decoding path. Furthermore, the left and right child node of $\nu$ are recursively decoded using the proposed decoder as specified in lines 13-16 of Algorithm~\ref{algo:p-FHT-FSCL}. Also note that a permutation sampled from the full symmetry group transforms a RM code to another RM code of similar length and order, whose frozen bit indicies are in general different from the frozen-bit indices of the original RM code. Therefore, one needs to re-permute the estimated codeword of the permuted LLR vector $\bm{\alpha}_{\nu_l}$ to reconstruct the original codeword \cite{geiselhart2020automorphism}. These operations are described in lines 18-21 of Algorithm~\ref{algo:p-FHT-FSCL}. Finally, the $\texttt{FHTL}(\cdot)$ and $\texttt{SPCL}(\cdot)$ functions are queried to decode the first-order RM subcodes and the SPC subcodes, respectively. The $\texttt{SPCL}(\cdot)$ function carries out the FSCL decoding operations as described in Section~\ref{sec:polar:FSCL}, while the details of the $\texttt{FHTL}(\cdot)$ function are provided in Algorithm~\ref{algo:FHTL}.}
	
	\begin{algorithm}[t]
		\DontPrintSemicolon
		\caption{$\texttt{FHTL}(\cdot)$ Decoding}
		\label{algo:FHTL}
		\SetKwInOut{Input}{Input}
		\SetKwInOut{Output}{Output}
		\SetKwInput{kwIn}{in}
		\SetKwFunction{FMain}{FHTL}
		\SetKwProg{Fn}{Function}{:}{}
		
		\Input{$\{\bm{\alpha}_{\nu_l}, \textup{PM}_l\}_{0 \leq l < L}$}		
		\Output{$\{\hat{\bm{x}}_{\nu_l}, \textup{PM}_l\}_{0 \leq l < L}$}
		\Fn{$\textup{\texttt{FHTL}}\left(\{\hat{\bm{x}}_{\nu_l}, \textup{PM}_l\}_{0 \leq l < L}\right)$}{
			\tcc{Perform FHT decoding for each  active decoding path}
			$\bm{\mathcal{E}} \leftarrow \emptyset$\\
			\For{$l \leftarrow 0$ \KwTo $L-1$}
			{
				$\{\bm{Q}_0,\ldots,\bm{Q}_{\min(L,2^s)-1}\} \leftarrow \texttt{FHT}(\bm{\alpha}_{\nu_l}, \textup{PM}_l)$\\
				$\bm{\mathcal{E}} \leftarrow \bm{\mathcal{E}} \cup \{\bm{Q}_0,\ldots,\bm{Q}_{\min(L,2^s)-1}\}$\\		
			}
			$\{\bm{Q}^*_0,\ldots,\bm{Q}^*_{L-1}\} \leftarrow \texttt{Sort}(\bm{\mathcal{E}})$\\
			\tcc{Return the $L$ best decoding paths}
			\For{$l \leftarrow 0$ \KwTo $L-1$}
			{
				$\hat{\bm{x}}_{\nu_l}\leftarrow\bm{Q}^*\{\hat{\bm{x}}_\nu\}$; 
				$\textup{PM}_l\leftarrow\bm{Q}^*\{\textup{PM}_l\}$\\
			}
			\Return{$\{\hat{\bm{x}}_{\nu_l}, \textup{PM}_l\}_{0 \leq l < L}$}
		}
	\end{algorithm}
	
	\textcolor{black}{In Algorithm~\ref{algo:FHTL}, for each input path with index $l$, we apply a modified FHT decoding algorithm on $\bm{\alpha}_{\nu_l}$ and generate the $\min(L,2^s)$ most probable decoding paths and their associated path metrics originated from $\bm{\alpha}_{\nu_l}$. The modified FHT decoding algorithm, $\texttt{FHT}(\cdot)$, that utilizes a low-complexity path metric computation scheme is provided in Algorithm~\ref{algo:FHTD}.} Specifically, after the FHT operations are applied to $\bm{\alpha}_{\nu_l}$ in Algorithm~\ref{algo:FHTD}, the indices of the largest absolute values of the transformed LLR vector $\bm{\alpha}^{\textup{FHT}}_{\nu_l}$ are obtained using a sorting algorithm (line 10 of Algorithm~\ref{algo:FHTD}). We only need to construct a maximum of $L$ best decoding paths generated from the current active path $l$ under FHT decoding. Therefore, it is not necessary to sort all the elements of the transformed LLR values $\abs{\bm{\alpha}^{\textup{FHT}}_{\nu_l}[i]}$ $(0 \leq i < N_\nu)$ given a small list size $L$. Consequently, the complexity of the sorting algorithm used in line 10 of Algorithm~\ref{algo:FHTD} is $\min(L2^s, s2^s)$. Note that $L2^s$ is the number of comparisons required by a straight-forward sorting algorithm that loops through the vector $\abs{\bm{\alpha}^{\textup{FHT}}_{\nu_l}}$ $L$ times to identify the indices of $L$ maximum elements, while a maximum of $s2^s$ comparisons are required by the merge sort algorithm, which is efficient for a large value of $L$ \cite{cormen2009introduction}. The sorting algorithm in line 10 of Algorithm~\ref{algo:FHTD} outputs the sorted indices $\{i^*_0, \ldots, i^*_{\min(L,2^s)-1}\}$, where $\abs{\bm{\alpha}^{\textup{FHT}}_{\nu_l}[i^*_0]} \geq \ldots \geq \abs{\bm{\alpha}^{\textup{FHT}}_{\nu_l}[i^*_{\min(L,2^s)-1}]}$, and $\bm{\alpha}^{\textup{FHT}}_{\nu_l}[i]$ indicates the $i$-th element of $\bm{\bm{\alpha}^{\textup{FHT}}_{\nu_l}}$. The indices $\{i^*_0, \ldots, i^*_{\min(L,2^s)-1}\}$ are then used to estimate the message word $\hat{\bm{u}}_\nu$ associated with $\bm{\alpha}_{\nu_l}$ (lines 12-17 of Algorithm~\ref{algo:FHTD}). $\text{dec2bin}(i)$ is a function that converts the decimal value of a bit index $i$ to its binary expansion represented by $s$ binary numbers.
		
	\begin{algorithm}[t]
		\DontPrintSemicolon
		\caption{$\texttt{FHT}(\cdot)$ Decoding}
		\label{algo:FHTD}
		\SetKwInOut{Input}{Input}
		\SetKwInOut{Output}{Output}
		\SetKwInput{kwIn}{in}	
		
		\Input{$\bm{\alpha}_{\nu}, \textup{PM}_l$}
		\Output{$\{\bm{Q}_0,\ldots,\bm{Q}_{\min(L,2^s)-1}\}$}
		\SetKwFunction{FMain}{FHT}
		\SetKwProg{Fn}{Function}{:}{}
		\Fn{\FMain{$\bm{\alpha}_{\nu}, \textup{PM}_l, l$}}{		
			\tcc{Initialization}
			$\bm{\alpha}^{\textup{FHT}}_{\nu_l} \leftarrow \bm{\alpha}_{\nu_l}$; 
			$\textup{LLR}_\textup{abs}\leftarrow \sum_{k=0}^{N_\nu-1} \abs{\bm{\alpha}_{\nu_l}[i]}$\\
			\tcc{Fast Hadamard Transform of $\bm{\alpha}_{\nu_l}$}
			\For{$t \leftarrow 0$ \KwTo $s-1$}
			{
				\For{$j \leftarrow 0$ \KwTo $2^{t+1} - 1$}
				{
					\For{$i \leftarrow j2^{s-t}$ \KwTo $j2^{s-t} + 2^{s-t-1} - 1$}
					{
						$a \leftarrow \bm{\alpha}^{\textup{FHT}}_{\nu_l}[i+2^{s-t-1}] - \bm{\alpha}^{\textup{FHT}}_{\nu_l}[i]$\\
						$b \leftarrow \bm{\alpha}^{\textup{FHT}}_{\nu_l}[i+2^{s-t-1}] + \bm{\alpha}^{\textup{FHT}}_{\nu_l}[i]$\\
						$\bm{\alpha}^{\textup{FHT}}_{\nu_l}[i] \leftarrow a$\\
						$\bm{\alpha}^{\textup{FHT}}_{\nu_l}[i+2^{s-t-1}] \leftarrow b$\\
					}
				}
			}
			
			\tcc{Obtain up to $L$ best decoding paths}
			$\{i^*_0, \ldots, i^*_{\min(L,2^s)-1}\} \leftarrow \texttt{Sort}\left(\abs{\bm{\alpha}^{\textup{FHT}}_{\nu_l}}\right)$\\
			
			\For{$j \leftarrow 0$ \KwTo $\min(L,2^s)-1$}
			{
				\tcc{Form the estimated messageword $\hat{\bm{u}}_\nu$}
				$m_s = \frac{1-\sgn{\bm{\alpha}^{\textup{FHT}}_{\nu_l}[i_j^*]}}{2}$\\
				$\{m_{s-1}, \ldots, m_0\} = \text{dec2bin}(2^s-1-i_j^*)$\\
				$t \leftarrow 0; \hat{\bm{u}}_\nu\leftarrow \bm{0}$\\
				\For{$k \leftarrow 0$ \KwTo $2^s-1$}
				{
					\If{$k$ \textup{is an information bit index}}{
						$\hat{\bm{u}}_\nu[k] \leftarrow m_t$;
						$ t \leftarrow t + 1$\\
					}		
				}
				$\hat{\bm{x}}_\nu \leftarrow \hat{\bm{u}}_\nu\bm{G}^{\otimes s}$\\
				$\textup{PM}_\nu \leftarrow \textup{PM}_l + \frac{1}{2}\left(\textup{LLR}_\textup{abs} -\abs{{\alpha}^{\textup{FHT}}_{\nu_l}[i_j^*]}\right)$\\
				\tcc{Form the output data structure}
				$\bm{Q}_j \leftarrow \{\hat{\bm{x}}_\nu, \textup{PM}_\nu\}$\\
			}
			
			\Return{$\{\bm{Q}_0,\ldots,\bm{Q}_{\min(L,2^s)-1}\}$}\\
		}
	\end{algorithm}
	
	\textcolor{black}{In line 18 of Algorithm~\ref{algo:FHTD}, $\hat{\bm{x}}_\nu$ is the estimated codeword corresponding to the $i^*_j$-th element of $\abs{\bm{\alpha}^{\textup{FHT}}_{\nu_l}}$. The path metric associated with $\hat{\bm{x}}_\nu$ is calculated as \cite{Ali_FSSCL}
	\begin{equation}
		\begin{split}
		\textup{PM}_\nu &= \textup{PM}_l+ \frac{1}{2}\sum_{k=0}^{N_\nu-1} \left(\abs{\bm{\alpha}_{\nu_l}[k]}-\left(1-2\hat{\bm{x}}_\nu[k]\right)\bm{\alpha}_{\nu_l}[k]\right),
		\end{split}
		\label{equ:PM:long}
	\end{equation}
	which can be rewritten as
	\begin{equation}
		\begin{split}
		\textup{PM}_\nu 
		&= \textup{PM}_l+ \frac{1}{2}\left[\text{LLR}_\text{abs} - \sum_{k=0}^{N_\nu-1} \left(1-2\hat{\bm{x}}_\nu[k]\right)\bm{\alpha}_{\nu_l}[k] \right]\\
		&= \textup{PM}_l+ \frac{1}{2}\left( \text{LLR}_\text{abs} -\abs{\bm{\alpha}^{\textup{FHT}}_{\nu_l}[i_j^*]}\right).
		\end{split}
		\label{equ:PM:short}
	\end{equation}
	Line 19 of Algorithm~\ref{algo:FHTD} computes the path metric associated with $\hat{\bm{x}}_\nu$ using (\ref{equ:PM:short}), which reuses the transformed LLR vector $\bm{\alpha}^{\textup{FHT}}_{\nu_l}$ to reduce the number of additions compared to (\ref{equ:PM:long}).} Algorithm~\ref{algo:FHTD} outputs the most probable decoding paths under FHT decoding as a set of the data structure $\bm{Q}_j$, i.e., $\{\bm{Q}_0,\ldots,\bm{Q}_{\min(L,2^s)-1}\}$, where $\bm{Q}_j$ consists of an estimated codeword $\hat{\bm{x}}_\nu$ and its corresponding path metric $\textup{PM}_\nu$. Note that as a maximum of $L$ best decoding paths are selected to continue the decoding after $\nu$ is visited, it is sufficient for Algorithm~\ref{algo:FHTD} to generate a maximum of $L$ candidate paths associated with each LLR vector $\bm{\alpha}_{\nu_l}$ $(0 \leq l < L)$.

	\begin{figure}[t]
		\centering	
		\begin{subfigure}{0.1\linewidth}
		\end{subfigure}
		\begin{subfigure}{0.8\linewidth}
			\centering	
			\input{Figures/example_FSC_simple.tikz.tex}\\
			\vspace*{0.5\baselineskip}
			\input{Figures/example_FSC_legend.tikz.tex}
		\end{subfigure}
		\caption{An example of the p-FHT-FSCL decoder with list size $L\ge1$ when applied to $\mathcal{RM}(3,5)$.}
		\label{fig:example}
	\end{figure}
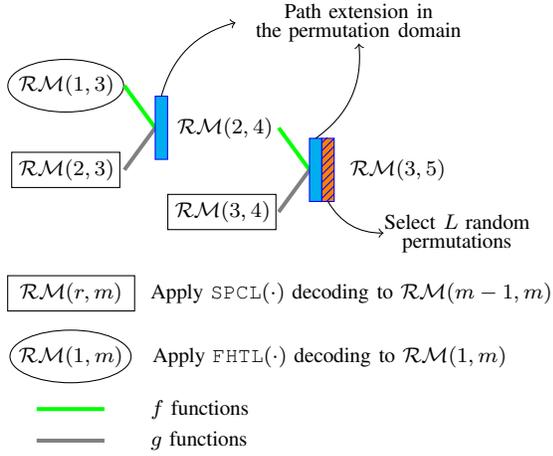
				
	\textcolor{black}{In Algorithm~\ref{algo:FHTL}, the outputs of the $\texttt{FHT}(\cdot)$ function that is applied to all the current active paths are stored in a set $\bm{\mathcal{E}}$. The sorting function applied to $\bm{\mathcal{E}}$ (line 6 of Algorithm~\ref{algo:FHTL}) generates a set of sorted data structures $\{\bm{Q}^*_0,\ldots,\bm{Q}^*_{L-1}\}$ such that $\bm{Q}^*_0\{\textup{PM}_\nu\} \leq \ldots \leq \bm{Q}^*_{L-1}\{\textup{PM}_\nu\}$, where $\bm{Q}^*_i\{\textup{PM}_\nu\}$ indicates the path metric associated with the $i$-th data structure $\bm{Q}^*_i$. We use the merge sort algorithm to output the sorted data structure $\bm{Q}^*_i$. Since the maximum size of $\bm{\mathcal{E}}$ is $L^2$, the maximum number of floating-point comparisons required by the sorting operation in line 6 of Algorithm~\ref{algo:FHTL} is $2L^2\log_2{L}$ \cite{cormen2009introduction}. The remainder of Algorithm~\ref{algo:FHTL} outputs the estimated codewords $\hat{\bm{x}}_{\nu_l}$ and the associated path metrics $\textup{PM}_l$ of all the best $L$ decoding paths. Fig.~\ref{fig:example} depicts an example of the proposed p-FHT-FSCL decoder on $\mathcal{RM}(3,5)$, where the proposed permutation decoding is only applied to $\mathcal{RM}(3,5)$ and its descendant $\mathcal{RM}(2,4)$. On the other hand, $\mathcal{RM}(3,4)$ and $\mathcal{RM}(2,3)$ are decoded using the FSCL decoding operations specified for the SPC nodes, while $\mathcal{RM}(1,3)$ is decoded using the FHTL decoding algorithm specified in Algorithm~\ref{algo:FHTL}.}
		
	\textcolor{black}{It can be observed in Algorithm~\ref{algo:p-FHT-FSCL} that the proposed p-FHT-FSCL decoder utilizes different subsets of the codeword permutations during the course of decoding. Therefore, to further utilize the rich symmetry group of RM codes, we run $M$ $(M>1)$ p-FHT-FSCL decoders with list size $L$ $(L\ge1)$ in parallel. Then, we select the output codeword that has the smallest path metric as the final estimated codeword. This improved decoder is denoted at p-FHT-FSCL-$L$-$M$. In Algorithm~\ref{algo:final}, we summarize the p-FHT-FSCL-$L$-$M$ decoder that utilizes path splitting in both codeword permutation and information bit domains.}
		
	\begin{algorithm}[t]
		\DontPrintSemicolon
		\caption{$\texttt{p-FHT-FSCL-}L\texttt{-}M(\cdot)$ Decoding}
		\label{algo:final}
		\SetKwInOut{Input}{Input}
		\SetKwInOut{Output}{Output}
		\SetKwInput{kwIn}{in}
		\SetKwFunction{SPRLD}{SP-RLD}
		
		\Input{$\bm{y}, M, L$}
		\Output{$\hat{\bm{x}}$}

		$\textup{PM}^*\leftarrow \infty$\\
		\For{$i\leftarrow 0$ \KwTo $M-1$}{
			\tcp{Initialization and decoding of the target RM code for each decoding attempt}
			$\{\bm{\alpha}_{\nu_l}\leftarrow \bm{y}, \textup{PM}_l \leftarrow 0\}_{0 \leq l < L}$\\
			$\{\hat{\bm{x}}_{\nu_l}, \textup{PM}_l\}_{0 \leq l < L} \leftarrow \texttt{p-FHT-FSCL}\left(\{\bm{\alpha}_{\nu_l}, \textup{PM}_l\}_{0 \leq l < L}\right)$\\
			$l^*\leftarrow \argmin\{\textup{PM}_0,\ldots,\textup{PM}_{L-1}\}$\\
			\tcp{Select the best estimated codeword from $M$ decoding attempts}
			\If{$\textup{PM}_{l^*}<\textup{PM}^*$}{
				$\textup{PM}^* \leftarrow \textup{PM}_{l^*}$\\
				$\hat{\bm{x}} \leftarrow \hat{\bm{x}}_{\nu_{l^*}}$
			}
		}
		\Return $\hat{\bm{x}}$
	\end{algorithm}
	
	\subsection{Performance Evaluation}
	
	\subsubsection{Quantitative Complexity Analysis}
	
	\textcolor{black}{We calculate the computational complexity of all the decoders presented in this paper by counting the number of floating-point additions and comparisons performed during the course of decoding for a received channel LLR vector $\bm{y}$. We summarize the computational complexities of all the decoding functions applied to a RM subcode of the proposed decoders in Table~\ref{tab:comp:general:1} and Table~\ref{tab:comp:general:2}. Furthermore, we compute the decoding latency of all the decoders presented in this paper by using the assumptions considered in \cite{Ali_FSSCL, Ardakani_TCOM}. Specifically, the hard decisions obtained from the LLR values and binary operations are computed instantaneously, and all the independent computations are calculated in parallel. Finally, we consider the number of time steps required by a merge sort algorithm to sort a vector of size $N$ to be $\log_2 {N}$ \cite{cormen2009introduction}.}
	
	\begin{table*}[t]
		\caption{Normalized computational complexities of different decoding functions required by the p-FHT-FSCL-$L$ decoder with $L>1$. The decoding functions are applied to a RM sub-code $\mathcal{RM}(r,m)$ visited by the decoding algorithm.}
		\def\arraystretch{1.25}
		\centering
		\begin{tabular}{c|cc|cc|c}	
			\toprule
			\multirow{2}{*}{Function} & \multicolumn{2}{c|}{Computation} & \multicolumn{2}{c|}{Sorting} & \multirow{2}{*}{Total}\\
			& LLR & Path Metric & LLR & Path Metric &\\
			\midrule
			$f(\cdot)$ & $L2^{m-1}$ &-&-&-&$L2^{m-1}$\\
			$g(\cdot)$ & $L2^{m-1}$ &-&-&-&$L2^{m-1}$\\
			$\mathcal{RM}(1,m)$ &$Lm2^m$&$L^2 2^m$& $\min(Lm2^m,L^2 2^m)$ &$2L^2\log_2L$&\textcolor{black}{$Lm2^m+\min(Lm2^m,L^2 2^m)+2L^2(2^{m-1}+\log_2 L)$}\\
			$\mathcal{RM}(m-1,m)$ &-&$L(1+2\tau)$&$\min(Lm2^m,L^2 2^m)$&$2\tau L(1+\log_2 L)$&\textcolor{black}{$\min(Lm2^m,L^2 2^m) + L\left[2\tau(2+\log_2L) + 1 \right]$}\\
		\end{tabular}
		\label{tab:comp:general:1}
	\end{table*}

	\begin{table}[t]
		\caption{Normalized computational complexities of different decoding functions of the p-FHT-FSCL-$1$ decoder. The decoding functions are applied to a RM sub-code $\mathcal{RM}(r,m)$ visited by the decoding algorithm.}
		\def\arraystretch{1.25}
		\centering
		\begin{tabular}{c | c | c | c}	
			\toprule
			Function & LLR Computation & LLR Sorting & Total\\
			\midrule
			$f(\cdot)$ & $2^{m-1}$ & - & $2^{m-1}$\\
			$g(\cdot)$ & $2^{m-1}$ & - & $2^{m-1}$\\
			$\mathcal{RM}(1,m)$ & $m2^m$&$2^m$& $2^m(m+1)$\\
			$\mathcal{RM}(m-1,m)$ &-&$2^m$&$2^m$\\
		\end{tabular}
		\label{tab:comp:general:2}
	\end{table}	
	
	\textcolor{black}{The FHT used in Algorithm~\ref{algo:FHTD} only uses in-place computations that do not require extra memory for the LLR values \cite{FHTD}. In addition, the path extension in the permutation domain of the proposed decoders is carried out sequentially for each decoding path. This allows the proposed decoders to maintain a similar memory requirement to store the LLR values compared to FHT-FSCL decoding with the same list size. The memory requirements in terms of the number of bits for the proposed decoders are summarized in Table~\ref{tab:mem:general}.}	
		
	\subsubsection{Comparison with FSCL and FHT-FSCL Decoding}
	
	\textcolor{black}{Fig.~\ref{fig:FER:1} illustrates the FER performance of FHT-FSCL and p-FHT-FSCL decoders with various list sizes $L \ge 1$. The FER performance of FSCL decoding with list size $32$ (FSCL-$32$) is also plotted for comparison. Furthermore, Fig.~\ref{fig:comp:1} plots the computational complexity $\mathcal{C}$, decoding latency in time steps $\mathcal{T}$, and memory requirement $\mathcal{M}$ in Kilobytes (KBs) of FHT-FSCL and p-FHT-FSCL decoders considered in Fig.~\ref{fig:FER:1}. Note that FHT-FSCL-$1$ indicates the FHT-FSC decoder and p-FHT-FSCL-$1$ indicates the proposed decoder that performs the path extension in the permutation domain until the first constituent RM code of order $1$ is visited, at which point the decoding operations are performed exactly similar to those of FHT-FSC decoding.}
	
	\textcolor{black}{It can be observed from Fig.~\ref{fig:FER:1} and Fig.~\ref{fig:comp:1} that by utilizing the proposed permutation decoding scheme, p-FHT-FSCL decoding significantly outperforms FHT-FSCL decoding with a similar list size $L>1$ for all the considered RM codes, while relatively maintaining all the complexity metrics. In particular, for $\mathcal{RM}(4,9)$, p-FHT-FSCL-$32$ provides an error-correction performance gain of $1$ dB at the target FER of $10^{-4}$ in comparison with FHT-FSCL-$32$, while having overheads of $9\%$ in the computation complexity and $2.4\%$ in the decoding latency. Note that p-FHT-FSCL-$32$ preserves the memory requirement of FHT-FSCL-$32$.}
	
	\begin{table}[t]
		\caption{Summary of the memory requirements of the proposed decoders.}
		\def\arraystretch{1.25}
		\centering
		\begin{tabular}{c | c}	
			\toprule
			Algorithm & Memory Requirement in Bits\\
			\midrule
			p-FHT-FSC-$1$ & $(2N+1)Q + N$\\
			\midrule
			\makecell{p-FHT-FSC-$1$-$M$\\$(M>0)$} & $(N+M(N+1))Q + MN$\\			
			\midrule
			\makecell{p-FHT-FSC-$L$\\$(L>1)$} & $N(L+1)Q+2LQ+ 2NL$\\
			\midrule
			\makecell{p-FHT-FSC-$L$-$M$\\$(L>1, M>1)$} & $N(LM+1)Q+ 2MLQ+ 2MNL$
		\end{tabular}
		\label{tab:mem:general}
	\end{table}
	
	{
		\begin{figure*}
			\centering		
			\input{./Figures/ferRM512_46_SSC_per.tikz.tex}
			\input{./Figures/ferRM512_130_SSC_per.tikz.tex}
			\input{./Figures/ferRM512_256_SSC_per.tikz.tex}\\	
			\ref{perf-legend-FSCL-FHT-9-2}	
			\caption{FER performance of the FHT-FSCL and p-FHT-FSC decoders for various RM codes. The FER values of the FSCL decoder with list size $32$ (FSCL-$32$) are also plotted for comparison.}
			\label{fig:FER:1}
		\end{figure*}
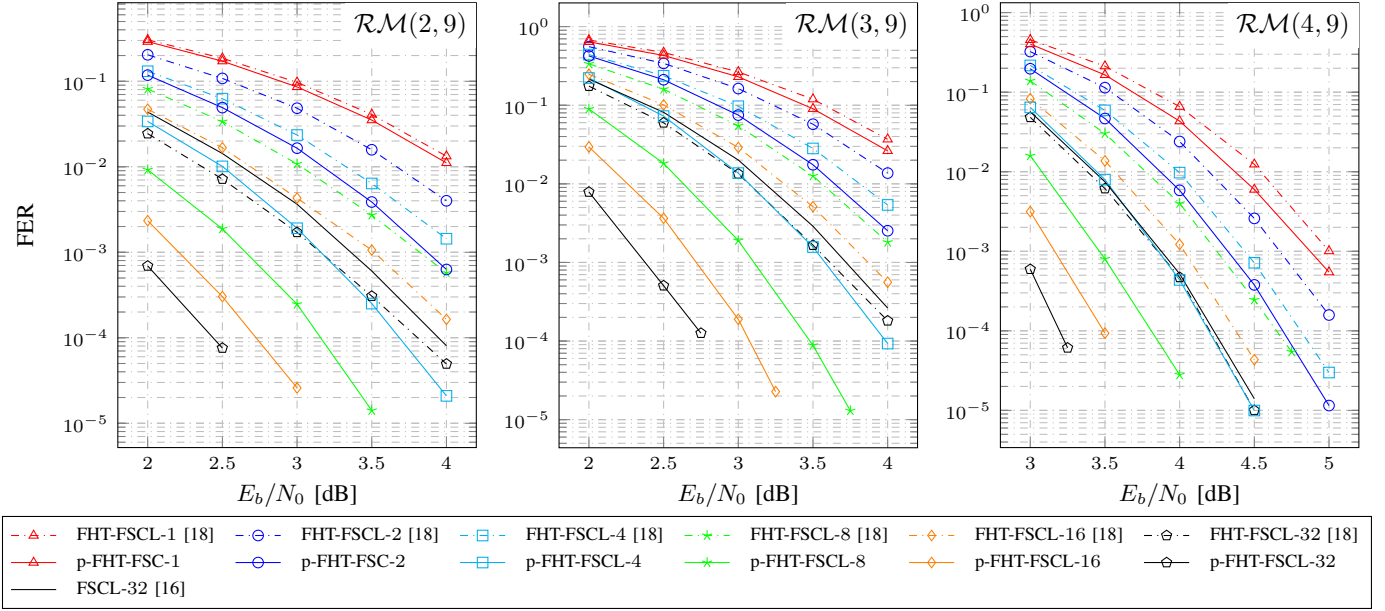
		
		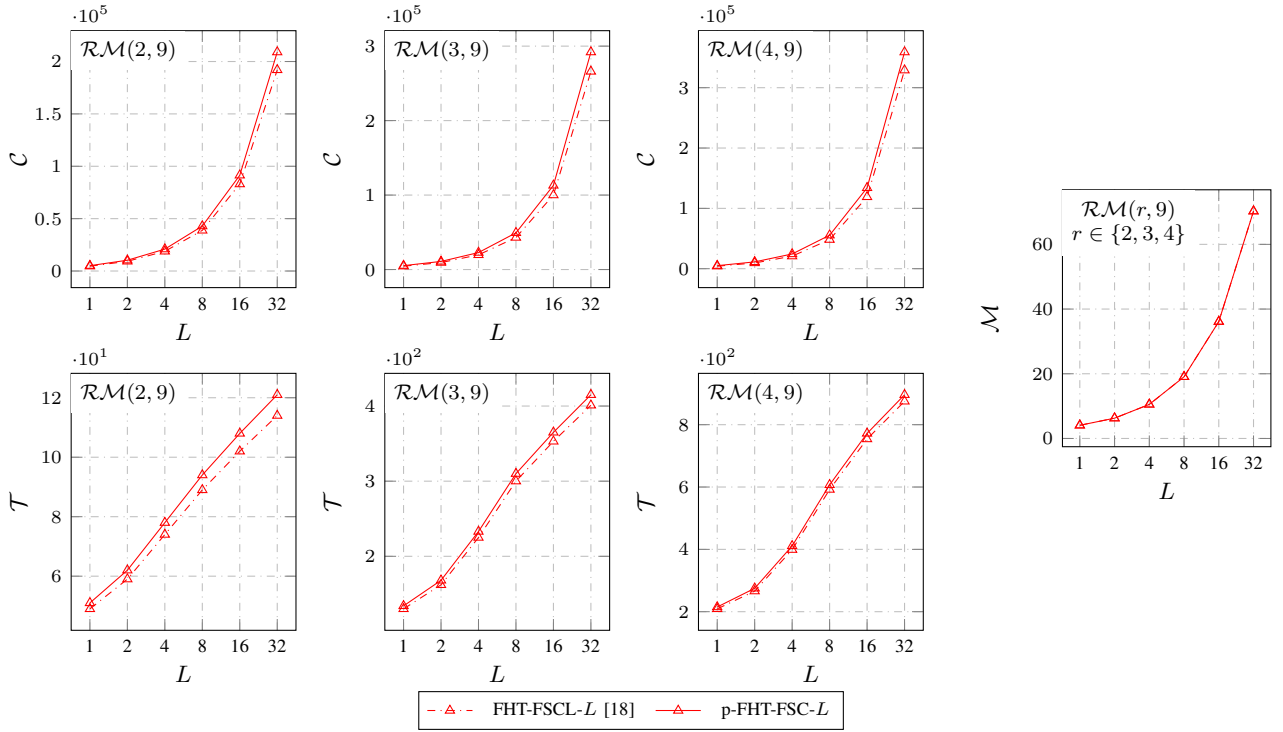
\begin{figure*}
			\centering
			\begin{subfigure}{0.72\linewidth}
				\centering
				\input{./Figures/comp_N512.tikz.tex}
				\input{./Figures/comp_N512_130.tikz.tex}
				\input{./Figures/comp_N512_256.tikz.tex}\\
				\input{./Figures/lat_N512.tikz.tex}
				\input{./Figures/lat_N512_130.tikz.tex}
				\input{./Figures/lat_N512_256.tikz.tex}\\
			\end{subfigure}
			\begin{subfigure}{0.24\linewidth}
				\centering
				\input{./Figures/mem_N512.tikz.tex}
			\end{subfigure}
			{\ref{comp-legend-FSCL-FHT-9-2}}
			\caption{Computational complexity $(\mathcal{C})$, decoding latency in time steps $(\mathcal{T})$, and memory requirement in KBs $(\mathcal{M})$ of FHT-FSCL-$L$ and p-FHT-FSCL-$L$ considered in Fig.~\ref{fig:FER:1}.}
			\label{fig:comp:1}
		\end{figure*}
		
		\begin{table*}
			\centering
			\setlength{\tabcolsep}{4pt}
			\renewcommand{\arraystretch}{1.25}
			\caption{Comparison of normalized computational complexity $(\mathcal{C})$, decoding latency in time steps $(\mathcal{T})$, and memory requirement in KBs $(\mathcal{M})$ of FSCL-$32$ \cite{Ali_FSSCL}, FHT-FSCL-$32$ \cite{Ghaddar20}, and p-FHT-FSCL-$4$, whose FER values are shown in Fig.~\ref{fig:FER:1}.}
			\footnotesize
			\begin{tabular}{c|ccc|ccc|ccc}
				\toprule
				&\multicolumn{3}{c|}{FSCL-$32$ \cite{Ali_FSSCL}}&\multicolumn{3}{c|}{FHT-FSCL-$32$ \cite{Ghaddar20}}&\multicolumn{3}{c}{p-FHT-FSCL-$4$}\\
				
				&\multicolumn{1}{c}{$\mathcal{C}$}&\multicolumn{1}{c}{$\mathcal{T}$}&\multicolumn{1}{c|}{$\mathcal{M}$}&\multicolumn{1}{c}{$\mathcal{C}$}&\multicolumn{1}{c}{$\mathcal{T}$}&\multicolumn{1}{c|}{$\mathcal{M}$}&\multicolumn{1}{c}{$\mathcal{C}$}&\multicolumn{1}{c}{$\mathcal{T}$}&\multicolumn{1}{c}{$\mathcal{M}$}\\
				
				\midrule
				$\mathcal{RM}(2,9)$&9.59E+04&373&70.25&1.92E+05&114&70.25&2.09E+04&78&10.53\\
				$\mathcal{RM}(3,9)$&1.55E+05&1039&70.25&2.66E+05&401&70.25&2.28E+04&233&10.53\\
				$\mathcal{RM}(4,9)$&2.24E+05&1991&70.25&3.29E+05&875&70.25&2.44E+04&411&10.53\\
			\end{tabular}
			\label{tab:comp:1}
		\end{table*}
	}	
		
	\textcolor{black}{Table~\ref{tab:comp:1} summarizes the computational complexity, decoding latency, and memory requirement of FSCL-$32$, FHT-FSCL-$32$, and p-FHT-FSCL-$4$, whose FER values are relatively similar at the target FER of $10^{-3}$ as shown in Fig.~\ref{fig:FER:1}. It can be observed from Table~\ref{tab:comp:1} that the negligible error-correction performance improvement of FHT-FSCL-$32$ with respect to FSCL-$32$ comes at the cost of significant computational complexity overhead, which is mainly caused by the sorting operations required by the FHT-based decoding algorithm. Note that the computational complexity required by the sorting operations under FHT-FSCL-based decoding increases significantly as the list size increases. On the other hand, by utilizing the proposed permutation decoding algorithm, the permuted FHT-FSCL decoder only requires a list size of $4$ to obtain a similar or better error-correction performance compared to FSCL-$32$ and FHT-FSCL-$32$ at the target FER of $10^{-3}$. The use of a much smaller list size ($4$ instead of $32$) also enables p-FHT-FSCL-$4$ to obtain significantly lower complexity metrics compared to FSCL-$32$ and FHT-FSCL-$32$ as observed from Table~\ref{tab:comp:1}. For example, in comparison with FHT-FSCL-$32$ for $\mathcal{RM}(4,9)$, p-FHT-FSCL-$4$ reduces $93\%$ of the computational complexity, $53\%$ of the decoding latency, and $85\%$ of the memory requirement.}
	
	{
	\begin{figure*}[t!]
		\centering
		\begin{subfigure}{0.49\linewidth}
			\centering
			\input{./Figures/ferRM256_37_FSC_per_rpa.tikz.tex}
		\end{subfigure}
		\begin{subfigure}{0.49\linewidth}
			\centering
			\input{./Figures/ferRM256_93_FSC_per_rpa.tikz.tex}
		\end{subfigure}\\
		\begin{subfigure}{0.49\linewidth}
			\centering
			\input{./Figures/ferRM256_163_FSC_per_rpa.tikz.tex}
		\end{subfigure}
		\begin{subfigure}{0.49\linewidth}
			\centering
			\input{./Figures/ferRM512_46_SSC_per_rpa.tikz.tex}
		\end{subfigure}\\
		\vspace*{10pt}
		\ref{perf-legend-all-8-2}	
		\caption{Error-correction performance of various RM decoders.}
		\label{fig:FER:2}
	\end{figure*}
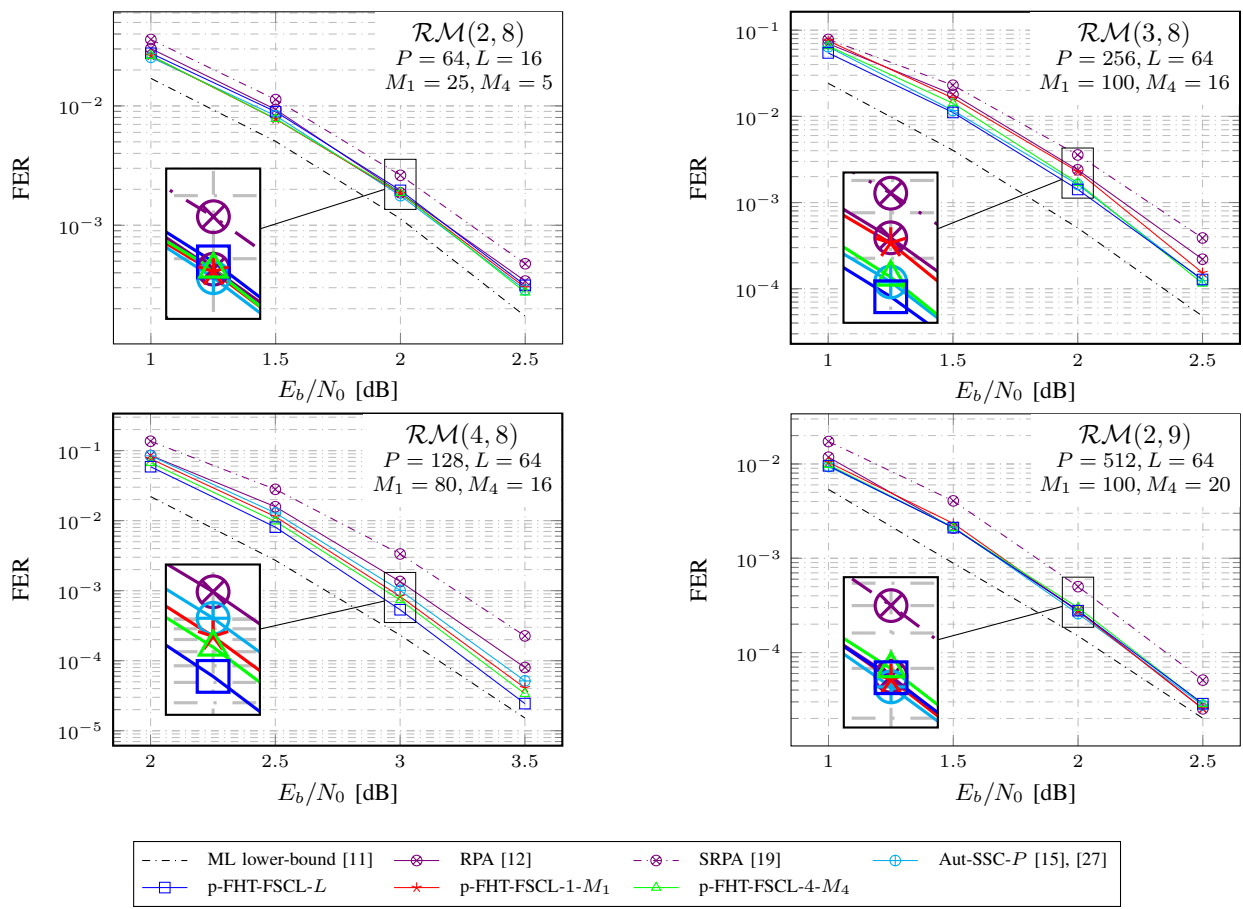
	
	\begin{table*}
		\centering
		\setlength{\tabcolsep}{1.8pt}
		\renewcommand{\arraystretch}{1.25}
		\caption{Comparison of normalized computational complexity $(\mathcal{C})$, decoding latency in time steps $(\mathcal{T})$, and memory requirement in KBs $(\mathcal{M})$ of various RM decoders considered in Fig.~\ref{fig:FER:2}.}
		\label{tab:comp:2}
		\footnotesize
		\begin{tabular}{c|ccc|ccc|cccc|cccc|cccc|cccc}
			\toprule
			&\multicolumn{3}{c|}{RPA \cite{Ye20}}&\multicolumn{3}{c|}{SRPA \cite{fathollahi2020sparse}}&\multicolumn{4}{c|}{Aut-SSC-$P$ \cite{gabi_fast_pcd, geiselhart2020automorphism}}&\multicolumn{4}{c|}{p-FHT-FSCL-$L$}&\multicolumn{4}{c|}{p-FHT-FSCL-$1$-$M_1$}&\multicolumn{4}{c}{p-FHT-FSCL-$4$-$M_4$}\\
			
			&\multicolumn{1}{c}{$\mathcal{C}$}&\multicolumn{1}{c}{$\mathcal{T}$}&\multicolumn{1}{c|}{$\mathcal{M}$}&\multicolumn{1}{c}{$\mathcal{C}$}&\multicolumn{1}{c}{$\mathcal{T}$}&\multicolumn{1}{c|}{$\mathcal{M}$}&\multicolumn{1}{c}{${P}$}&\multicolumn{1}{c}{$\mathcal{C}$}&\multicolumn{1}{c}{$\mathcal{T}$}&\multicolumn{1}{c|}{$\mathcal{M}$}&\multicolumn{1}{c}{${L}$}&\multicolumn{1}{c}{$\mathcal{C}$}&\multicolumn{1}{c}{$\mathcal{T}$}&\multicolumn{1}{c|}{$\mathcal{M}$}&\multicolumn{1}{c}{${M_1}$}&\multicolumn{1}{c}{$\mathcal{C}$}&\multicolumn{1}{c}{$\mathcal{T}$}&\multicolumn{1}{c|}{$\mathcal{M}$}&\multicolumn{1}{c}{$M_4$}&\multicolumn{1}{c}{$\mathcal{C}$}&\multicolumn{1}{c}{$\mathcal{T}$}&\multicolumn{1}{c}{$\mathcal{M}$}\\
						
			\midrule
			$\mathcal{RM}(2,8)$&1.8E+6&3592&135.5&6.5E+5&3592&69.2&64&9.4E+4&80&67.0&16&4.5E+4&92&18.1&25&5.7E+4&46&26.9&5&4.8E+4&68&22.4\\
			$\mathcal{RM}(3,8)$&4.3E+8&6184&556.8&7.9E+7&6184&281.5&256&4.5E+5&147&265.0&64&5.2E+5&356&69.5&100&2.3E+5&100&104.5&16&1.7E+5&173&69.5\\
			$\mathcal{RM}(4,8)$&3.8E+10&7816&922.2&3.6E+9&7816&465.2&128&2.2E+5&165&133.0&64&5.7E+5&673&69.5&80&1.7E+5&131&83.8&16&1.8E+5&251&69.5\\
			$\mathcal{RM}(2,9)$&9.8E+6&10250&535.1&3.4E+6&10250&271.6&512&1.5E+6&106&1058.0&64&5.4E+5&134&138.5&100&5.1E+5&58&208.6&20&4.2E+5&83&172.6\\
		\end{tabular}
	\end{table*}
	}
		
	\subsubsection{Comparison with Permuted SC-Based Decoding and RPA-Based Decoding}
	
	\textcolor{black}{Fig.~\ref{fig:FER:2} illustrates the error-correction performance of the simplified SC (SSC) \cite{gabi_fast_pcd} decoder when utilizing $P$ random codeword permutations sampled from the full symmetry group of the codes (Aut-SSC-$P$) and that of the RPA \cite{Ye20} and SRPA \cite{fathollahi2020sparse} decoders. In addition, we consider the following configurations of the proposed decoders in Fig.~\ref{fig:FER:2}: p-FHT-FSCL-$L$, p-FHT-FSCL-$1$-$M_1$, and p-FHT-FSCL-$4$-$M_4$. Note that p-FHT-FSCL-$1$-$M_1$ runs $M_1$ p-FHT-FSCL-$1$ decoders in parallel while p-FHT-FSCL-$4$-$M_4$ runs $M_4$ p-FHT-FSCL-$4$ decoders in parallel. Also note that under p-FHT-FSCL-$1$, only the path extension in the permutation domain is carried out. The values of $L$, $M_1$, and $M_4$ are selected to provide a similar error-correction performance of the proposed decoders in comparison with RPA and Aut-SSC decoding at the target FER of $10^{-3}$. We also plot the empirical ML lower bounds of the FER values for all the RM codes considered in Fig.~\ref{fig:FER:2} \cite{Dumer06}. In Table~\ref{tab:comp:2}, we summarize the computational complexity, decoding latency in time steps, and memory requirement in KBs of all the decoders considered in Fig.~\ref{fig:FER:2}.}
	
	\textcolor{black}{It can be observed from Fig.~\ref{fig:FER:2} and Table~\ref{tab:comp:2} that all the permutation decoding algorithms outperform the RPA and SRPA decoders in various complexity metrics while having a similar or better error-correction performance compared to the RPA decoder. Note that p-FHT-FSCL-$L$ is the most memory-efficient decoding algorithm, while p-FHT-FSCL-$1$-$M_1$ provides the lowest decoding latency in time steps among all the decoders. On the other hand, the p-FHT-FSCL-$4$-$M_4$ configuration enables a better decoding latency and memory requirement trade-off compared to p-FHT-FSCL-$L$ and p-FHT-FSCL-$1$-$M_1$ settings and obtains the smallest computational complexity for $\mathcal{RM}(3,8)$ and $\mathcal{RM}(2,9)$. In addition, for $\mathcal{RM}(2,8)$ and $\mathcal{RM}(3,8)$, p-FHT-FSCL-$4$-$M_4$ has a similar memory requirement to p-FHT-FSCL-$L$, while having significantly smaller computational complexity and decoding latency. Compared to Aut-SSC-$512$ for $\mathcal{RM}(2,9)$ and at a target FER of $10^{-4}$, p-FHT-FSCL-$4$-$M_4$ reduces $72\%$ of the computational complexity, $22\%$ of the decoding latency, and $84\%$ of the memory requirement. Compared to SRPA decoding for $\mathcal{RM}(2,9)$ and at the target FER of $10^{-4}$, p-FHT-FSCL-$4$-$M_4$ provides $36\%$ reduction in the memory consumption and $88\%$ reduction in the computational complexity, while achieving several order-of-magnitude lower decoding latency and $0.14$ dB gain in error-correction performance.}
		
	\section{Conclusion}
	\label{sec:conclud}
	
	\textcolor{black}{In this paper, we introduced a novel permutation decoding algorithm for {Reed-Muller} (RM) codes tailored to the existing fast successive-cancellation list decoder with fast Hadamard transform (FHT-FSCL). The proposed permuted FHT-FSCL (p-FHT-FSCL) decoder performs the path extension in the codeword permutation domain to select the $L$ best decoding paths until the first constituent RM code of order $1$ is decoded. As the p-FHT-FSCL decoder utilizes different subsets of the codeword permutations sampled from the full symmetry group of the codes, the error-correction performance of RM codes can be significantly improved by running $M$ p-FHT-FSCL decoders with list size $L$ in parallel. We performed a detailed numerical performance analysis of the computational complexity, decoding latency, and memory requirement of the proposed decoders and compared with those of sparse recursive-projection aggregation, FHT-FSCL, and the state-of-the-art permuted successive-cancellation (Aut-SSC) decoders. The simulation results show that for the RM code of length $512$ with order $2$, the proposed decoder with $L=4$ and $M=20$ reduces $72\%$ of the computational complexity, $22\%$ of the decoding latency, and $84\%$ of the memory requirement with respect to the state-of-the-art Aut-SSC decoder with $512$ random codeword permutations, while obtaining a similar error-correction performance at the target frame error rate of $10^{-4}$.}
	

\end{document}

%% file: Figures/PolarFactorGraph.tikz
\begin{tikzpicture}[scale=1.0]

\def\N{16}
\def\xM{7.5}
\def\xss{\xM/\N}
\def\xs{\xss/1}
\def\ys{0.35}
\def\gain{1.0}
\def\markSize{2}
\def\PEmarkSize{3.5}

\draw[dashed] (0*\xs,0*\ys) -- (0*\xs,7.75*\ys*\gain) node[above]{\footnotesize{$s_0$}};
\draw[dashed] (2*\xs,0*\ys) -- (2*\xs,7.75*\ys*\gain) node[above]{\footnotesize{$s_1$}};
\draw[dashed] (5*\xs,0*\ys) -- (5*\xs,7.75*\ys*\gain) node[above]{\footnotesize{$s_2$}};
\draw[dashed] (10*\xs,0*\ys) -- (10*\xs,7.75*\ys*\gain) node[above]{\footnotesize{$s_3$}};

\foreach \i in{0,...,7}
{
	\pgfmathsetmacro\bIndex{int(7-\i)};
	\draw[] (0,\i*\ys) -- (10*\xs,\i*\ys);
	
	\draw[] plot[mark=*, mark size = \markSize, mark options={fill=white}] coordinates {(0*\xs,\i*\ys)};
	
	\ifthenelse{\i = 4}{\draw[] plot[mark=*, mark size = \markSize, mark options={fill=black}] coordinates {(0*\xs,\i*\ys)};}
	
	\ifthenelse{\i = 8}{\draw[] plot[mark=*, mark size = \markSize, mark options={fill=black}] coordinates {(0*\xs,\i*\ys)};}
		
	\ifthenelse{\i < 3}{\draw[] plot[mark=*, mark size = \markSize, mark options={fill=black}] coordinates {(0*\xs,\i*\ys)};}
	
	\draw[] plot[mark=*, mark size = \markSize, mark options={color=gray}] coordinates {(2*\xs,\i*\ys)};
	\draw[] plot[mark=*, mark size = \markSize, mark options={color=gray}] coordinates {(5*\xs,\i*\ys)};
	\draw[] plot[mark=*, mark size = \markSize, mark options={color=gray}] coordinates {(10*\xs,\i*\ys)};
	
	\node[text width=0.5cm] at (-0.5*\xs,\i*\ys) {\footnotesize{$u_{\bIndex}$}};
	\node[text width=0.5cm] at (11*\xs,\i*\ys) {\footnotesize{$x_{\bIndex}$}};
}

\draw[] plot[mark=oplus, mark size = \PEmarkSize, mark options={fill=white}] coordinates {(\xs,\ys)} --
plot[mark=*, mark size = \PEmarkSize, mark options={fill=white}] coordinates {(\xs,0)} node[above=-0.2]{\scriptsize{=}};

\draw[] plot[mark=oplus, mark size = \PEmarkSize, mark options={fill=white}] coordinates {(\xs,3*\ys)} --
plot[mark=*, mark size = \PEmarkSize, mark options={fill=white}] coordinates {(\xs,2*\ys)} node[above=-0.2]{\scriptsize{=}};

\draw[] plot[mark=oplus, mark size = \PEmarkSize, mark options={fill=white}] coordinates {(\xs,5*\ys)} --
plot[mark=*, mark size = \PEmarkSize, mark options={fill=white}] coordinates {(\xs,4*\ys)} node[above=-0.2]{\scriptsize{=}};

\draw[] plot[mark=oplus, mark size = \PEmarkSize, mark options={fill=white}] coordinates {(\xs,7*\ys)} --
plot[mark=*, mark size = \PEmarkSize, mark options={fill=white}] coordinates {(\xs,6*\ys)} node[above=-0.2]{\scriptsize{=}};

\draw[] plot[mark=oplus, mark size = \PEmarkSize, mark options={fill=white}] coordinates {(3*\xs,3*\ys)} --
plot[mark=*, mark size = \PEmarkSize, mark options={fill=white}] coordinates {(3*\xs,\ys)} node[above=-0.2]{\scriptsize{=}};

\draw[] plot[mark=oplus, mark size = \PEmarkSize, mark options={fill=white}] coordinates {(3*\xs,7*\ys)} --
plot[mark=*, mark size = \PEmarkSize, mark options={fill=white}] coordinates {(3*\xs,5*\ys)} node[above=-0.2]{\scriptsize{=}};

\draw[] plot[mark=oplus, mark size = \PEmarkSize, mark options={fill=white}] coordinates {(4*\xs,2*\ys)} --
plot[mark=*, mark size = \PEmarkSize, mark options={fill=white}] coordinates {(4*\xs,0)} node[above=-0.2]{\scriptsize{=}};

\draw[] plot[mark=oplus, mark size = \PEmarkSize, mark options={fill=white}] coordinates {(4*\xs,6*\ys)} --
plot[mark=*, mark size = \PEmarkSize, mark options={fill=white}] coordinates {(4*\xs,4*\ys)} node[above=-0.2]{\scriptsize{=}};

\draw[] plot[mark=oplus, mark size = \PEmarkSize, mark options={fill=white}] coordinates {(9*\xs,4*\ys)} --
plot[mark=*, mark size = \PEmarkSize, mark options={fill=white}] coordinates {(9*\xs,0*\ys)} node[above=-0.2]{\scriptsize{=}};

\draw[] plot[mark=oplus, mark size = \PEmarkSize, mark options={fill=white}] coordinates {(8*\xs,5*\ys)} --
plot[mark=*, mark size = \PEmarkSize, mark options={fill=white}] coordinates {(8*\xs,1*\ys)} node[above=-0.2]{\scriptsize{=}};

\draw[] plot[mark=oplus, mark size = \PEmarkSize, mark options={fill=white}] coordinates {(7*\xs,6*\ys)} --
plot[mark=*, mark size = \PEmarkSize, mark options={fill=white}] coordinates {(7*\xs,2*\ys)} node[above=-0.2]{\scriptsize{=}};

\draw[] plot[mark=oplus, mark size = \PEmarkSize, mark options={fill=white}] coordinates {(6*\xs,7*\ys)} --
plot[mark=*, mark size = \PEmarkSize, mark options={fill=white}] coordinates {(6*\xs,3*\ys)} node[above=-0.2]{\scriptsize{=}};



\end{tikzpicture}

%% file: Figures/SCPE.tikz
\begin{tikzpicture}[scale=0.7]
\def\markSize{3}
\draw[] plot[mark=*, mark size = \markSize, mark options={color=gray}] coordinates {(0,0.75)} node[left=0.1] {$\alpha_{s,i},\beta_{s,i}$} -- plot[mark=*, mark size = \markSize, mark options={color=gray}] coordinates {(2,0.75)} node[right=0.1] {$\alpha_{s+1,i},\beta_{s+1,i}$};

\draw[] plot[mark=*, mark size = \markSize, mark options={color=gray}] coordinates {(0,0)} node[left=0.1] {$\alpha_{s,i+2^s},\beta_{s,i+2^s}$} -- plot[mark=*, mark size = \markSize, mark options={color=gray}] coordinates {(2,0)} node[right=0.1] {$\alpha_{s+1,i+2^s},\beta_{s+1,i+2^s}$};

\draw[] plot[mark=oplus, mark size = 6.5, mark options={fill=white}] coordinates {(1,0.75)} --
plot[mark=*, mark size = 6.5, mark options={fill=white}] coordinates {(1,0)} node[above=-0.225]{=};
\end{tikzpicture}

%% file: Figures/RLD.tikz.tex
\usetikzlibrary{arrows, decorations}

\tikzstyle{vecArrow} = [thick, decoration={markings,mark=at position
	1 with {\arrow[semithick]{open triangle 60}}},
double distance=1.4pt, shorten >= 5.5pt,
preaction = {decorate},
postaction = {draw,line width=1.4pt, white,shorten >= 4.5pt}]
\tikzstyle{innerWhite} = [semithick, white,line width=1.4pt, shorten >= 4.5pt]

\begin{tikzpicture}[scale=1.0]
\footnotesize
\def\N{16}
\def\xM{7.5}
\def\xss{\xM/\N}
\def\xs{\xss/1}
\def\ys{0.4}
\def\Ygain{1.075}
\def\Xgain{0.8}
\def\markSize{2.5}
\def\PEmarkSize{3.8}


\draw[] (19*\xs*\Xgain,3.5*\ys) -- (17*\xs*\Xgain,1.5*\ys);
\draw[] (19*\xs*\Xgain,3.5*\ys) -- (17*\xs*\Xgain,5.5*\ys);
\draw[] plot[mark=*, mark size = \markSize, mark options={fill=gray}] coordinates {(19*\xs*\Xgain,3.5*\ys)};

\draw[] plot[mark=square*, mark size = \markSize, mark options={fill=black}] coordinates {(17*\xs*\Xgain,1.5*\ys)};

\draw[] plot[mark=triangle*, mark size = \markSize*1.25, mark options={fill=black}] coordinates {(17*\xs*\Xgain,4*\ys+1.5*\ys)};

\draw[] (15*\xs*\Xgain,3*\ys) -- (17*\xs*\Xgain,1.5*\ys);
\draw[] plot[mark=triangle*, mark size = 1.25*\markSize, mark options={fill=white}] coordinates {(15*\xs*\Xgain,3*\ys)};

\draw[] (15*\xs*\Xgain,0*\ys) -- (17*\xs*\Xgain,1.5*\ys);
\draw[] plot[mark=diamond*, mark size = 1.25*\markSize, mark options={fill=white}] coordinates {(15*\xs*\Xgain,0*\ys)};

\node[text width=0cm] at (13*\xs*\Xgain,4*\ys) {\footnotesize{$\mathcal{RM}(0,1)$}};
\node[text width=0cm] at (13*\xs*\Xgain,-1*\ys) {\footnotesize{$\mathcal{RM}(1,1)$}};

\node[text width=0cm] at (17.5*\xs*\Xgain,6*\ys) {\footnotesize{$\mathcal{RM}(0,2)$}};
\node[text width=0cm] at (17.5*\xs*\Xgain,1*\ys) {\footnotesize{$\mathcal{RM}(1,2)$}};
\node[text width=0cm] at (19.5*\xs*\Xgain,3.4*\ys) {\footnotesize{$\mathcal{RM}(1,3)$}};
\end{tikzpicture}

%% file: Figures/FSCL.tikz.tex
\usetikzlibrary{arrows, decorations}

\tikzstyle{vecArrow} = [thick, decoration={markings,mark=at position
	1 with {\arrow[semithick]{open triangle 60}}},
double distance=1.4pt, shorten >= 5.5pt,
preaction = {decorate},
postaction = {draw,line width=1.4pt, white,shorten >= 4.5pt}]
\tikzstyle{innerWhite} = [semithick, white,line width=1.4pt, shorten >= 4.5pt]

\begin{tikzpicture}[scale=1.0]
\footnotesize
\def\N{16}
\def\xM{7.5}
\def\xss{\xM/\N}
\def\xs{\xss/1}
\def\ys{0.4}
\def\Ygain{1.075}
\def\Xgain{0.8}
\def\markSize{2.5}
\def\PEmarkSize{3.8}


\draw[] (19*\xs*\Xgain,3.5*\ys) -- (17*\xs*\Xgain,1*\ys);
\draw[] (19*\xs*\Xgain,3.5*\ys) -- (17*\xs*\Xgain,6*\ys);
\draw[] plot[mark=*, mark size = \markSize, mark options={fill=gray}] coordinates {(19*\xs*\Xgain,3.5*\ys)};

\draw[] plot[mark=square*, mark size = \markSize, mark options={fill=black}] coordinates {(17*\xs*\Xgain,1*\ys)};

\draw[] plot[mark=triangle*, mark size = \markSize*1.25, mark options={fill=black}] coordinates {(17*\xs*\Xgain,6*\ys)};

\node[text width=0cm] at (15*\xs*\Xgain,6.8*\ys) {\footnotesize{$\mathcal{RM}(0,2)$}};
\node[text width=0cm] at (15*\xs*\Xgain,-0.2*\ys) {\footnotesize{$\mathcal{RM}(1,2)$}};
\node[text width=0cm] at (19*\xs*\Xgain,2.5*\ys) {\footnotesize{$\mathcal{RM}(1,3)$}};

\end{tikzpicture}

%% file: Figures/projection.tikz.tex
\usetikzlibrary{matrix,decorations.pathreplacing, calc, positioning,fit}

\begin{tikzpicture}[
	>=stealth,thick,baseline,
	every left delimiter/.style={xshift=.75em},
	every right delimiter/.style={xshift=-.75em},
	scale=1]

	\matrix [matrix of math nodes, left delimiter={[},right delimiter={]}](A) at (0, 0) { 
		0 & 1\\
		2 & 3\\
		4 & 5\\
		6 & 7\\
	} ;

	\matrix [matrix of math nodes, left delimiter={[},right delimiter={]}](B) at (1.25, 0) 
	{ 
		0 & 2\\
		1 & 3\\
		4 & 6\\
		5 & 7\\
	} ;

	\matrix [matrix of math nodes, left delimiter={[},right delimiter={]}](C) at (2.5, 0) 
	{ 
		0 & 3\\
		1 & 2\\
		4 & 7\\
		5 & 6\\
	} ;

		\matrix [matrix of math nodes, left delimiter={[},right delimiter={]}](D) at (3.75, 0) 
	{ 
		0 & 4\\
		1 & 5\\
		2 & 6\\
		3 & 7\\
	} ;

		\matrix [matrix of math nodes, left delimiter={[},right delimiter={]}](E) at (5, 0) 
	{ 
		0 & 5\\
		1 & 4\\
		2 & 7\\
		3 & 6\\
	} ;

		\matrix [matrix of math nodes, left delimiter={[},right delimiter={]}](F) at (6.25, 0) 
	{ 
		0 & 6\\
		1 & 7\\
		2 & 4\\
		3 & 5\\
	} ;

		\matrix [matrix of math nodes, left delimiter={[},right delimiter={]}](G) at (7.5, 0) 
	{ 
		0 & 7\\
		1 & 6\\
		2 & 5\\
		3 & 4\\
	} ;

	\node[text width=1cm] at (0.2,1.35) {\footnotesize{$j=0$}};
	\node[text width=1cm] at (1.4,1.35) {\footnotesize{$j=1$}};
	\node[text width=1cm] at (2.6,1.35) {\footnotesize{$j=2$}};
	\node[text width=1cm] at (3.9,1.35) {\footnotesize{$j=3$}};		
	\node[text width=1cm] at (5.1,1.35) {\footnotesize{$j=4$}};	
	\node[text width=1cm] at (6.4,1.35) {\footnotesize{$j=5$}};
	\node[text width=1cm] at (7.6,1.35) {\footnotesize{$j=6$}};		
	
	\draw[rounded corners, dashdotted] (0.75, 0.5) rectangle (1.75, 0.9) {};	
	\draw [->, dashdotted] (1.25, 0.5) -- (1.25, -1.25);	
	\node[text width=3cm] at (2, -2) {$k=0$\\ $\mathbb{B}_{0,0,0}=0$\\ $\mathbb{B}_{0,0,1}=2$};
	
	\draw[rounded corners, dashdotted] (4.5, -0.0) rectangle (5.5, -0.425) {};	
	\draw [->, dashdotted] (5, -0.425) -- (5, -1.25);	
	\node[text width=3cm] at (5.75, -2) {$k=2$\\ $\mathbb{B}_{4,2,0}=2$\\ $\mathbb{B}_{4,2,1}=7$};

\end{tikzpicture}

%% file: Figures/example_FSC_simple.tikz.tex
\usetikzlibrary{patterns}
\usetikzlibrary{arrows, decorations, patterns}

\tikzstyle{vecArrow} = [thick, decoration={markings,mark=at position
	1 with {\arrow[semithick]{open triangle 60}}},
double distance=1.4pt, shorten >= 5.5pt,
preaction = {decorate},
postaction = {draw,line width=1.4pt, white,shorten >= 4.5pt}]
\tikzstyle{innerWhite} = [semithick, white,line width=1.4pt, shorten >= 4.5pt]

\begin{tikzpicture}[scale=0.7]
\footnotesize	
\def\N{16}
\def\xM{7.5}
\def\xss{\xM/\N}
\def\xs{\xss/1}
\def\ys{0.4}
\def\Ygain{1.075}
\def\Xgain{0.8}
\def\markSize{3}
\def\PEmarkSize{3.8}

\draw (6.125*\xs,2*\ys) circle [x radius=1.1, y radius=0.5, rotate=0];	
\node[text width=1.25cm, align=center] at (6.125*\xs,2*\ys) {\footnotesize{$\mathcal{RM}(1,3)$}};
\node[text width=1.25cm, draw=black, align=center] at (6.125*\xs,-2*\ys) {\footnotesize{$\mathcal{RM}(2,3)$}};

\draw[line width=1.5pt, color=green] (8.5*\xs,2*\ys)--(9.75*\xs,0*\ys);
\draw[line width=1.5pt, color=gray] (8.5*\xs,-2*\ys)--(9.75*\xs,0*\ys);
\draw[preaction={fill=cyan}, draw=blue] (9.75*\xs,-1.5*\ys) rectangle (10.25*\xs,1.5*\ys);
\node[text width=1.22cm] at (12.5*\xs,0*\ys) {\footnotesize{$\mathcal{RM}(2, 4)$}};
\node[text width=1.25cm, draw=black] at (12.45*\xs,-4*\ys) {\footnotesize{$\mathcal{RM}(3, 4)$}};

\draw[line width=1.5pt, color=green] (14.75*\xs,0*\ys)--(16*\xs,-2*\ys);
\draw[line width=1.5pt, color=gray] (14.75*\xs,-4*\ys)--(16*\xs,-2*\ys);
\draw[preaction={fill=cyan}, draw=blue] (16*\xs,-3.5*\ys) rectangle (16.5*\xs,-0.5*\ys);
\node[text width=1.22cm] at (19.5*\xs,-2*\ys) {\footnotesize{$\mathcal{RM}(3, 5)$}};

\draw[preaction={fill=orange}, pattern=north east lines, pattern color=blue, draw=blue] (16.5*\xs,-3.5*\ys) rectangle (17*\xs,-0.5*\ys);

\path[->] (10*\xs, 1.5*\ys) edge [bend left] (13*\xs, 5*\ys);
\path[->] (16.25*\xs, -0.5*\ys) edge [bend right] (18*\xs,4.0*\ys);
\node[text width=4cm, font=\footnotesize\linespread{0.8}\selectfont, align=center] at (18*\xs, 5*\ys) {Path extension in\\the permutation domain};

\path[->] (16.75*\xs,-3.5*\ys) edge [bend right] (19*\xs, -5*\ys);
\node[text width=6cm, font=\footnotesize\linespread{0.8}\selectfont, align=center] at (22*\xs,-5*\ys) {Select $L$ random\\permutations};

\end{tikzpicture}

%% file: Figures/example_FSC_legend.tikz.tex
\usetikzlibrary{patterns}
\usetikzlibrary{arrows, decorations, patterns}

\tikzstyle{vecArrow} = [thick, decoration={markings,mark=at position
	1 with {\arrow[semithick]{open triangle 60}}},
double distance=1.4pt, shorten >= 5.5pt,
preaction = {decorate},
postaction = {draw,line width=1.4pt, white,shorten >= 4.5pt}]
\tikzstyle{innerWhite} = [semithick, white,line width=1.4pt, shorten >= 4.5pt]

\begin{tikzpicture}[scale=0.7]
	\footnotesize	
	\def\N{16}
	\def\xM{7.5}
	\def\xss{\xM/\N}
	\def\xs{\xss/1}
	\def\ys{0.4}
	\def\Ygain{1.075}
	\def\Xgain{0.8}
	\def\markSize{3}
	\def\PEmarkSize{3.8}

	\node[text width=1.5cm, align=center, draw=black] at (6.15*\xs,-8*\ys) {\footnotesize{$\mathcal{RM}(r,m)$}};
	\node[text width=6.25 cm, align=left, font=\footnotesize] at (18.9*\xs,-8*\ys) {Apply \texttt{SPCL}$(\cdot)$ decoding to $\mathcal{RM}(m-1,m)$};
	
	\draw (6.125*\xs,-11*\ys) circle [x radius=1.15, y radius=0.5, rotate=0];
	\node[text width=1.5cm, align=center] at (6.15*\xs,-11*\ys) {\footnotesize{$\mathcal{RM}(1,m)$}};
	\node[text width=6 cm, align=left, font=\footnotesize] at (18.6*\xs,-11*\ys) {Apply \texttt{FHTL}$(\cdot)$ decoding to $\mathcal{RM}(1,m)$};
	
	\draw[line width=1.5pt, color=green] (4.75*\xs,-13.5*\ys)--(7.5*\xs,-13.5*\ys);
	\node[text width=5 cm, align=left, font=\footnotesize] at (17*\xs,-13.5*\ys) {$f$ functions};
	
	\draw[line width=1.5pt, color=gray] (4.75*\xs,-15*\ys)--(7.5*\xs,-15*\ys);
	\node[text width=5 cm, align=left, font=\footnotesize] at (17*\xs,-15*\ys) {$g$ functions};
	
\end{tikzpicture}

%% file: Figures/ferRM512_46_SSC_per.tikz.tex
\begin{tikzpicture}
	\pgfplotsset{
		label style = {font=\fontsize{9pt}{7.2}\selectfont},
		tick label style = {font=\fontsize{7pt}{7.2}\selectfont}
	}
	
	\begin{axis}[
		scale = 1,
		ymode=log,
		xlabel={$E_b/N_0$ [\text{dB}]}, xlabel style={yshift=0.5em},
		ytick={1e-7, 1e-6, 1e-5,1e-4,1e-3,1e-2,1e-1,1e0},
		xtick={2,2.5,3,3.5,4},
		ylabel={FER}, 
		grid=both,
		ymajorgrids=true,
		xmajorgrids=true,
		grid style=dashdotted,
		width=0.35\linewidth, height=7.5cm,
		thin,
		mark size=2.25,
		legend cell align=left,
		legend style={
			column sep= 2mm,
			font=\fontsize{7pt}{7.2}\selectfont,
		},
		legend to name=perf-legend-FSCL-FHT-9-2,
		legend columns=6,
		]

		\addplot[
		mark=triangle,
		thin,
		red,
		mark size=2,
		dashdotted,
		mark options={solid},
		]
		table {
			2	0.30563056
			2.5	1.86E-1
			3	0.09710971
			3.5	4.09E-2
			4	0.01330133
		};
		\addlegendentry{FHT-FSCL-$1$ \cite{Ghaddar20}}
		
		\addplot[
		color=blue,
		mark=o,
		thin,
		mark size=2,
		dashdotted,
		mark options={solid},
		]
		table {
			2	2.04E-01
			2.5	1.08E-01
			3	4.84E-02
			3.5	1.58E-02
			4	4.00E-03
		};
		\addlegendentry{FHT-FSCL-$2$ \cite{Ghaddar20}}
		
		\addplot[
		color=cyan,
		mark=square,
		thin,
		mark size=2,
		dashdotted,
		mark options={solid},
		]
		table {
			2	1.31E-01
			2.5	6.26E-02
			3	2.35E-02
			3.5	6.38E-03
			4	1.44E-03
		};
		\addlegendentry{FHT-FSCL-$4$ \cite{Ghaddar20}}
		
		\addplot[
		color=green,
		mark=star,
		thin,
		mark size=2,
		dashdotted,
		mark options={solid},
		]
		table {
			2	8.12E-02
			2.5	3.38E-02
			3	1.08E-02
			3.5	2.72E-03
			4	5.74E-04
		};
		\addlegendentry{FHT-FSCL-$8$ \cite{Ghaddar20}}

		\addplot[
		color=orange,
		mark=diamond,
		thin,
		mark size=2,
		dashdotted,
		mark options={solid},
		]
		table {
			2	4.71E-02
			2.5	1.67E-02
			3	4.30E-03
			3.5	1.06E-03
			4	1.64E-04
		};
		\addlegendentry{FHT-FSCL-$16$ \cite{Ghaddar20}}
		
		\addplot[
		color=black,
		mark=pentagon,
		thin,
		mark size=2,
		dashdotted,
		mark options={solid},
		]
		table {
			2	2.44E-02
			2.5	7.18E-03
			3	1.71E-03
			3.5	3.07E-04
			4	4.94E-05
		};
		\addlegendentry{FHT-FSCL-$32$ \cite{Ghaddar20}}
		
		\addplot[
			mark=triangle,
			thin,
			red,
			mark size=2,
			solid,
			mark options={solid},
		]
		table {
			2	2.91E-01
			2.5	1.72E-01
			3	8.59E-02
			3.5	3.52E-02
			4	1.11E-02
		};
		\addlegendentry{p-FHT-FSC-$1$}
		
		\addplot[
		color=blue,
		mark=o,
		thin,
		mark size=2,
		solid,
		mark options={solid},
		]
		table {
			2	1.18E-01
			2.5	4.92E-02
			3	1.65E-02
			3.5	3.87E-03
			4	6.29E-04
		};
		\addlegendentry{p-FHT-FSC-$2$}
		
		\addplot[
		color=cyan,
		mark=square,
		thin,
		mark size=2,
		solid,
		mark options={solid},
		]
		table {
			2	3.39E-02
			2.5	1.01E-02
			3	1.91E-03
			3.5	2.51E-04
			4.0	2.09E-05
		};
		\addlegendentry{p-FHT-FSCL-$4$}
		
		\addplot[
			color=green,
			mark=star,
			thin,
			mark size=2,
			solid,
			mark options={solid},
		]
		table {
			2	9.22E-03
			2.5	1.88E-03
			3	2.47E-04
			3.5	1.41E-05
		};
		\addlegendentry{p-FHT-FSCL-$8$}
		
		\addplot[
		color=orange,
		mark=diamond,
		thin,
		mark size=2,
		solid,
		mark options={solid},
		]
		table {
			2.0	2.34E-03
			2.5 3.03E-04
			3.0	2.59E-05			
		};
		\addlegendentry{p-FHT-FSCL-$16$}

		\addplot[
		color=black,
		mark=pentagon,
		thin,
		mark size=2,
		solid,
		mark options={solid},
		]
		table {
			2.0	6.92E-04
			2.5	7.58E-05
		};
		\addlegendentry{p-FHT-FSCL-$32$}
		
		\addplot[
			color=black,
			mark=none,
			thin,
			mark size=2,
			solid,
			mark options={solid},
		]
		table {
			2	4.37E-02
			2.5	1.43E-02
			3	3.65E-03
			3.5	6.04E-04
			4	8.06E-05
		};
		\addlegendentry{FSCL-$32$ \cite{Ali_FSSCL}}
		
		\node[anchor=north east, fill=white] at (rel axis cs:1,1) {{$\mathcal{RM}(2,9)$}};
	\end{axis}
\end{tikzpicture}

%% file: Figures/ferRM512_130_SSC_per.tikz.tex
\begin{tikzpicture}
	\pgfplotsset{
		label style = {font=\fontsize{9pt}{7.2}\selectfont},
		tick label style = {font=\fontsize{7pt}{7.2}\selectfont}
	}
	
	\begin{axis}[
		scale = 1,
		ymode=log,
		xlabel={$E_b/N_0$ [\text{dB}]}, xlabel style={yshift=0.5em},
		ytick={1e-7, 1e-6, 1e-5,1e-4,1e-3,1e-2,1e-1,1e-0},
		xtick={2,2.5,3,3.5,4,5,6},
		ylabel={}, 
		grid=both,
		ymajorgrids=true,
		xmajorgrids=true,
		grid style=dashdotted,
		width=0.35\linewidth, height=7.5cm,
		thin,
		mark size=2.25,
		legend cell align=left,
		legend style={
			column sep= 2mm,
			font=\fontsize{7pt}{7.2}\selectfont,
		},
		]
		\addplot[
			dashdotted,
			mark=triangle,
			thin,
			red,
			mark size=2,
			dashdotted,
			mark options={solid},
		]
		table {
			2	0.67646765
			2.5	0.46934693
			3	0.26552655
			3.5	0.11941194
			4	0.03690369
		};
		
		\addplot[
			mark=o,
			thin,
			blue,
			mark size=2,
			dashdotted,
			mark options={solid},
		]
		table {
			2	5.56E-01
			2.5	3.41E-01
			3	1.63E-01
			3.5	5.74E-02
			4	1.37E-02
		};
		
		\addplot[
			mark=square,
			thin,
			cyan,
			mark size=2,
			dashdotted,
			mark options={solid},
		]
		table {
			2	4.41E-01
			2.5	2.37E-01
			3	9.72E-02
			3.5	2.82E-02
			4	5.39E-03
		};
		
		\addplot[
			color=green,
			mark=star,
			thin,
			mark size=2,
			dashdotted,
			mark options={solid},
		]
		table {
			2	3.36E-01
			2.5	1.60E-01
			3	5.48E-02
			3.5	1.26E-02
			4	1.81E-03
		};
		
		\addplot[
			color=orange,
			mark=diamond,
			thin,
			mark size=2,
			dashdotted,
			mark options={solid},
		]
		table {
			2	2.48E-01
			2.5	1.01E-01
			3	2.90E-02
			3.5	5.12E-03
			4	5.60E-04
		};

		\addplot[
			color=black,
			mark=pentagon,
			thin,
			mark size=2,
			dashdotted,
			mark options={solid},
		]
		table {
			2	1.75E-01
			2.5	5.98E-02
			3	1.34E-02
			3.5	1.66E-03
			4	1.81E-04
		};
		
				\addplot[
		mark=triangle,
		thin,
		red,
		mark size=2,
		solid,
		mark options={solid},
		]
		table {
			2	6.41E-01
			2.5	4.30E-01
			3	2.30E-01
			3.5	9.02E-02
			4	2.63E-02
		};
		
		\addplot[
		color=blue,
		mark=o,
		thin,
		mark size=2,
		solid,
		mark options={solid},
		]
		table {
			2	4.24E-01
			2.5	2.11E-01
			3	7.49E-02
			3.5	1.75E-02
			4	2.53E-03
		};
		
		\addplot[
		color=cyan,
		mark=square,
		thin,
		mark size=2,
		solid,
		mark options={solid},
		]
		table {
			2	2.22E-01
			2.5	7.28E-02
			3	1.37E-02
			3.5	1.57E-03
			4	9.29E-05
		};
		
		\addplot[
		color=green,
		mark=star,
		thin,
		mark size=2,
		solid,
		mark options={solid},
		]
		table {
			2	8.97E-02
			2.5	1.82E-02
			3	1.90E-03
			3.5	8.82E-05
			3.75	1.30E-05
		};
		
		\addplot[
		color=orange,
		mark=diamond,
		thin,
		mark size=2,
		solid,
		mark options={solid},
		]
		table {
			2.0	2.95E-02
			2.5	3.65E-03
			3.0	1.89E-04
			3.25	2.27E-05
		};

		\addplot[
		color=black,
		mark=pentagon,
		thin,
		mark size=2,
		solid,
		mark options={solid},
		]
		table {
			2.0	7.88E-03
			2.5	5.06E-04
			2.75	1.26E-04
		};
		
						\addplot[
		color=black,
		mark=none,
		thin,
		mark size=2,
		solid,
		mark options={solid},
		]
		table {
			2	2.15E-01
			2.5	8.13E-02
			3	2.01E-02
			3.5	2.89E-03
			4.0	2.63E-04
		};
		
		\node[anchor=north east, fill=white] at (rel axis cs:1,1) {{$\mathcal{RM}(3,9)$}};
	\end{axis}
\end{tikzpicture}

%% file: Figures/ferRM512_256_SSC_per.tikz.tex
\begin{tikzpicture}
	\pgfplotsset{
		label style = {font=\fontsize{9pt}{7.2}\selectfont},
		tick label style = {font=\fontsize{7pt}{7.2}\selectfont}
	}
	
	\begin{axis}[
		scale = 1,
		ymode=log,
		xlabel={$E_b/N_0$ [\text{dB}]}, xlabel style={yshift=0.5em},
		ytick={1e-7, 1e-6, 1e-5,1e-4,1e-3,1e-2,1e-1,1e-0},
		xtick={2.5,3,3.5,4,4.5,5},
		ylabel={}, 
		grid=both,
		ymajorgrids=true,
		xmajorgrids=true,
		grid style=dashdotted,
		width=0.35\linewidth, height=7.5cm,
		thin,
		mark size=2.25,
		legend cell align=left,
		legend style={
			column sep= 2mm,
			font=\fontsize{7pt}{7.2}\selectfont,
		},
		]

		\addplot[
		mark=triangle,
		thin,
		red,
		mark size=2,
		mark options={solid},
		dashdotted,
		]
		table {
			3	0.45544554
			3.5	0.21242124
			4	0.06580658
			4.5	0.01240124
			5	0.00100625
		};
		
		\addplot[
			mark=o,
			thin,
			blue,
			mark size=2,
			dashdotted,
			mark options={solid},
		]
		table {
			3	3.24E-01
			3.5	1.14E-01
			4	2.40E-02
			4.5	2.59E-03
			5	1.58E-04
		};
		
		\addplot[
			mark=square,
			thin,
			cyan,
			mark size=2,
			dashdotted,
			mark options={solid},
		]
		table {
			3	2.17E-01
			3.5	5.91E-02
			4	9.74E-03
			4.5	7.17E-04
			5	2.99E-05
		};
		
		\addplot[
			color=green,
			mark=star,
			thin,
			mark size=2,
			dashdotted,
			mark options={solid},
		]
		table {
			3	1.39E-01
			3.5	3.02E-02
			4	4.00E-03
			4.5	2.45E-04
			4.75	5.44E-05
		};

		\addplot[
		color=orange,
		mark=diamond,
		thin,
		mark size=2,
		dashdotted,
		mark options={solid},
		]
		table {
			3	8.28E-02
			3.5	1.37E-02
			4	1.22E-03
			4.5	4.34E-05
		};
		
		\addplot[
			color=black,
			mark=pentagon,
			thin,
			mark size=2,
			dashdotted,
			mark options={solid},
		]
		table {
			3.0	4.83E-02
			3.5	6.14E-03
			4.0	4.70E-04
			4.5	1.00E-05
		};
		
		\addplot[
			mark=triangle,
			thin,
			red,
			mark size=2,
			solid,
			mark options={solid},
		]
		table {
			3	4.03E-01
			3.5	1.65E-01
			4	4.33E-02
			4.5	5.98E-03
			5	5.46E-04
		};
		
		\addplot[
		color=blue,
		mark=o,
		thin,
		mark size=2,
		solid,
		mark options={solid},
		]
		table {
			3	1.97E-01
			3.5	4.67E-02
			4	5.84E-03
			4.5	3.78E-04
			5.0	1.15E-05	
		};
		
		\addplot[
		color=cyan,
		mark=square,
		thin,
		mark size=2,
		solid,
		mark options={solid},
		]
		table {
			3	6.44E-02
			3.5	7.96E-03
			4	4.36E-04
			4.5	9.96E-06
		};
		
		\addplot[
		color=green,
		mark=star,
		thin,
		mark size=2,
		solid,
		mark options={solid},
		]
		table {
			3	1.59E-02
			3.5	8.00E-04
			4.0	2.80E-05
		};
		
		\addplot[
		color=orange,
		mark=diamond,
		thin,
		mark size=2,
		solid,
		mark options={solid},
		]
		table {
			3.0	3.15E-03
			3.5	9.35E-05
		};

		\addplot[
		color=black,
		mark=pentagon,
		thin,
		mark size=2,
		solid,
		mark options={solid},
		]
		table {
			3.0	5.95E-04
			3.25	6.11E-05
		};
		
								\addplot[
		color=black,
		mark=none,
		thin,
		mark size=2,
		solid,
		mark options={solid},
		]
		table {
			3	5.68E-02
			3.5	7.47E-03
			4.0	5.20E-04
			4.5	1.40E-05
		};
		
		\node[anchor=north east, fill=white] at (rel axis cs:1,1) {{$\mathcal{RM}(4,9)$}};
	\end{axis}
\end{tikzpicture}

%% file: Figures/comp_N512.tikz.tex
\begin{tikzpicture}
	\pgfplotsset{
		label style = {font=\fontsize{9pt}{7.2}\selectfont},
		tick label style = {font=\fontsize{7pt}{7.2}\selectfont}
	}
	
	\begin{axis}[
		scale = 1,
		xlabel={$L$}, xlabel style={yshift=0.5em},
		ytick={0,0.5e5,1e5,1.5e5,2e5},
		xtick={1,2,3,4,5,6},
		xticklabels={1,2,4,8,16,32},
		ylabel={$\mathcal{C}$}, ylabel style={yshift=-1.5em},
		grid=both,
		ymajorgrids=true,
		xmajorgrids=true,
		grid style=dashdotted,
		width=0.35\linewidth, height=5cm,
		thin,
		mark size=2.25,
		legend cell align=left,
		legend style={
			column sep= 2mm,
			font=\fontsize{7pt}{7.2}\selectfont,
		},
		legend to name=comp-legend-FSCL-FHT-9-2,
		legend columns=6,
		]

		\addplot[
		mark=triangle,
		thin,
		red,
		mark size=2,
		dashdotted,
		mark options={solid},
		]
		table {
			1 4.59E+03
			2 9.26E+03
			3 1.88E+04
			4 3.88E+04
			5 8.30E+04
			6 1.92E+05
		};
		\addlegendentry{FHT-FSCL-$L$ \cite{Ghaddar20}}
		
		\addplot[
			mark=triangle,
			thin,
			red,
			mark size=2,
			solid,
			mark options={solid},
		]
		table {
			1 5.11E+03
			2 1.03E+04
			3 2.09E+04
			4 4.29E+04
			5 9.14E+04
			6 2.09E+05
		};
		\addlegendentry{p-FHT-FSC-$L$}
		
		\node[anchor=north west, fill=white] at (rel axis cs:0,1) {\footnotesize{$\mathcal{RM}(2,9)$}};
	\end{axis}
\end{tikzpicture}

%% file: Figures/comp_N512_130.tikz.tex
\begin{tikzpicture}
	\pgfplotsset{
		label style = {font=\fontsize{9pt}{7.2}\selectfont},
		tick label style = {font=\fontsize{7pt}{7.2}\selectfont}
	}
	
	\begin{axis}[
		scale = 1,
		xlabel={$L$}, xlabel style={yshift=0.5em},
		xtick={1,2,3,4,5,6},
		xticklabels={1,2,4,8,16,32},
		ylabel={$\mathcal{C}$}, ylabel style={yshift=-1.5em},
		grid=both,
		ymajorgrids=true,
		xmajorgrids=true,
		grid style=dashdotted,
		width=0.35\linewidth, height=5cm,
		thin,
		mark size=2.25,
		legend cell align=left,
		legend style={
			column sep= 2mm,
			font=\fontsize{7pt}{7.2}\selectfont,
		},
		legend to name=comp-legend-FSCL-FHT-9-3,
		legend columns=6,
		]

		\addplot[
		mark=triangle,
		thin,
		red,
		mark size=2,
		dashdotted,
		mark options={solid},
		]
		table {
			1 4.48E+03
			2 9.37E+03
			3 1.97E+04
			4 4.30E+04
			5 1.00E+05
			6 2.66E+05
		};
		\addlegendentry{FHT-FSCL-$L$ \cite{Ghaddar20}}
		
		\addplot[
			mark=triangle,
			thin,
			red,
			mark size=2,
			solid,
			mark options={solid},
		]
		table {
			1 5.25E+03
			2 1.09E+04
			3 2.28E+04
			4 4.93E+04
			5 1.13E+05
			6 2.92E+05
		};
		\addlegendentry{p-FHT-FSC-$L$-1}
		
		\node[anchor=north west, fill=white] at (rel axis cs:0,1) {\footnotesize{$\mathcal{RM}(3,9)$}};
	\end{axis}
\end{tikzpicture}

%% file: Figures/comp_N512_256.tikz.tex
\begin{tikzpicture}
	\pgfplotsset{
		label style = {font=\fontsize{9pt}{7.2}\selectfont},
		tick label style = {font=\fontsize{7pt}{7.2}\selectfont}
	}
	
	\begin{axis}[
		scale = 1,
		xlabel={$L$}, xlabel style={yshift=0.5em},
		xtick={1,2,3,4,5,6},
		xticklabels={1,2,4,8,16,32},
		ylabel={$\mathcal{C}$}, ylabel style={yshift=-1.5em},
		grid=both,
		ymajorgrids=true,
		xmajorgrids=true,
		grid style=dashdotted,
		width=0.35\linewidth, height=5cm,
		thin,
		mark size=2.25,
		legend cell align=left,
		legend style={
			column sep= 2mm,
			font=\fontsize{7pt}{7.2}\selectfont,
		},
		legend to name=comp-legend-FSCL-FHT-9-4,
		legend columns=6,
		]

		\addplot[
		mark=triangle,
		thin,
		red,
		mark size=2,
		dashdotted,
		mark options={solid},
		]
		table {
			1 4.13E+03
			2 9.51E+03
			3 2.08E+04
			4 4.82E+04
			5 1.19E+05
			6 3.29E+05
		};
		\addlegendentry{FHT-FSCL-$L$ \cite{Ghaddar20}}
		
		\addplot[
			mark=triangle,
			thin,
			red,
			mark size=2,
			solid,
			mark options={solid},
		]
		table {
			1 5.03E+03
			2 1.13E+04
			3 2.44E+04
			4 5.55E+04
			5 1.34E+05
			6 3.59E+05
		};
		\addlegendentry{p-FHT-FSC-$L$-1}
		
		\node[anchor=north west, fill=white] at (rel axis cs:0,1) {\footnotesize{$\mathcal{RM}(4,9)$}};
	\end{axis}
\end{tikzpicture}

%% file: Figures/lat_N512.tikz.tex
\begin{tikzpicture}
	\pgfplotsset{
		label style = {font=\fontsize{9pt}{7.2}\selectfont},
		tick label style = {font=\fontsize{7pt}{7.2}\selectfont}
	}
	
	\begin{axis}[
		scale = 1,
		scaled y ticks=base 10:-1,
		xlabel={$L$}, xlabel style={yshift=0.5em},
		xtick={1,2,3,4,5,6},
		xticklabels={1,2,4,8,16,32},
		ylabel={$\mathcal{T}$}, ylabel style={yshift=-1.5em},
		grid=both,
		ymajorgrids=true,
		xmajorgrids=true,
		grid style=dashdotted,
		width=0.35\linewidth, height=5cm,
		thin,
		mark size=2.25,
		legend cell align=left,
		legend style={
			column sep= 2mm,
			font=\fontsize{7pt}{7.2}\selectfont,
		},
		legend to name=lat-legend-FSCL-FHT-9-2,
		legend columns=6,
		]

		\addplot[
		mark=triangle,
		thin,
		red,
		mark size=2,
		dashdotted,
		mark options={solid},
		]
		table {
			1 49
			2 59
			3 74
			4 89
			5 102
			6 114
		};
		\addlegendentry{FHT-FSCL-$L$ \cite{Ghaddar20}}
		
		\addplot[
			mark=triangle,
			thin,
			red,
			mark size=2,
			solid,
			mark options={solid},
		]
		table {
			1 51
			2 62
			3 78
			4 94
			5 108
			6 121
		};
		\addlegendentry{p-FHT-FSC-$L$}
		
		\node[anchor=north west, fill=white] at (rel axis cs:0,1) {\footnotesize{$\mathcal{RM}(2,9)$}};
	\end{axis}
\end{tikzpicture}

%% file: Figures/lat_N512_130.tikz.tex
\begin{tikzpicture}
	\pgfplotsset{
		label style = {font=\fontsize{9pt}{7.2}\selectfont},
		tick label style = {font=\fontsize{7pt}{7.2}\selectfont}
	}
	
	\begin{axis}[
		scale = 1,
		scaled y ticks=base 10:-2,
		xlabel={$L$}, xlabel style={yshift=0.5em},
		xtick={1,2,3,4,5,6},
		xticklabels={1,2,4,8,16,32},
		ylabel={$\mathcal{T}$}, ylabel style={yshift=-1.5em},
		grid=both,
		ymajorgrids=true,
		xmajorgrids=true,
		grid style=dashdotted,
		width=0.35\linewidth, height=5cm,
		thin,
		mark size=2.25,
		legend cell align=left,
		legend style={
			column sep= 2mm,
			font=\fontsize{7pt}{7.2}\selectfont,
		},
		legend to name=lat-legend-FSCL-FHT-9-3,
		legend columns=6,
		]

		\addplot[
		mark=triangle,
		thin,
		red,
		mark size=2,
		dashdotted,
		mark options={solid},
		]
		table {
			1 130
			2 162
			3 225
			4 300
			5 353
			6 401
		};
		\addlegendentry{FHT-FSCL-$L$ \cite{Ghaddar20}}
		
		\addplot[
			mark=triangle,
			thin,
			red,
			mark size=2,
			solid,
			mark options={solid},
		]
		table {
			1 134
			2 168
			3 233
			4 310
			5 365
			6 415
		};
		\addlegendentry{p-FHT-FSC-$L$-1}
		
		\node[anchor=north west, fill=white] at (rel axis cs:0,1) {\footnotesize{$\mathcal{RM}(3,9)$}};
	\end{axis}
\end{tikzpicture}

%% file: Figures/lat_N512_256.tikz.tex
\begin{tikzpicture}
	\pgfplotsset{
		label style = {font=\fontsize{9pt}{7.2}\selectfont},
		tick label style = {font=\fontsize{7pt}{7.2}\selectfont}
	}
	
	\begin{axis}[
		scale = 1,
		scaled y ticks=base 10:-2,
		xlabel={$L$}, xlabel style={yshift=0.5em},
		xtick={1,2,3,4,5,6},
		xticklabels={1,2,4,8,16,32},
		ylabel={$\mathcal{T}$}, ylabel style={yshift=-1.5em},
		grid=both,
		ymajorgrids=true,
		xmajorgrids=true,
		grid style=dashdotted,
		width=0.35\linewidth, height=5cm,
		thin,
		mark size=2.25,
		legend cell align=left,
		legend style={
			column sep= 2mm,
			font=\fontsize{7pt}{7.2}\selectfont,
		},
		legend to name=lat-legend-FSCL-FHT-9-4,
		legend columns=6,
		]

		\addplot[
		mark=triangle,
		thin,
		red,
		mark size=2,
		dashdotted,
		mark options={solid},
		]
		table {
			1 209
			2 266
			3 399
			4 592
			5 754
			6 875
		};
		\addlegendentry{FHT-FSCL-$L$ \cite{Ghaddar20}}
		
		\addplot[
			mark=triangle,
			thin,
			red,
			mark size=2,
			solid,
			mark options={solid},
		]
		table {
			1 215
			2 275
			3 411
			4 607
			5 772
			6 896
		};
		\addlegendentry{p-FHT-FSC-$L$-1}
		
		\node[anchor=north west, fill=white] at (rel axis cs:0,1) {\footnotesize{$\mathcal{RM}(4,9)$}};
	\end{axis}
\end{tikzpicture}

%% file: Figures/mem_N512.tikz.tex
\begin{tikzpicture}
	\pgfplotsset{
		label style = {font=\fontsize{9pt}{7.2}\selectfont},
		tick label style = {font=\fontsize{7pt}{7.2}\selectfont}
	}
	
	\begin{axis}[
		scale = 1,
		xlabel={$L$}, xlabel style={yshift=0.5em},
		xtick={1,2,3,4,5,6},
		xticklabels={1,2,4,8,16,32},
		ylabel={$\mathcal{M}$}, ylabel style={yshift=-0.75em},
		grid=both,
		ymajorgrids=true,
		xmajorgrids=true,
		grid style=dashdotted,
		width=1\linewidth, height=5cm,
		thin,
		mark size=2.25,
		legend cell align=left,
		legend style={
			column sep= 2mm,
			font=\fontsize{7pt}{7.2}\selectfont,
		},
		legend to name=comp-legend-FSCL-FHT-9-2,
		legend columns=6,
		]

		\addplot[
		mark=triangle,
		thin,
		red,
		mark size=2,
		dashdotted,
		mark options={solid},
		]
		table {
			1 4.07
			2 6.27
			3 10.53
			4 19.06
			5 36.12
			6 70.25
		};
		\addlegendentry{FHT-FSCL-$L$ \cite{Ghaddar20}}
		
		\addplot[
			mark=triangle,
			thin,
			red,
			mark size=2,
			solid,
			mark options={solid},
		]
		table {
			1 4.07
			2 6.27
			3 10.53
			4 19.06
			5 36.12
			6 70.25
		};
		\addlegendentry{p-FHT-FSC-$L$}
		
		\node[anchor=north west, fill=white] at (rel axis cs:0,1) {\footnotesize{\makecell{$\mathcal{RM}(r,9)$\\$r\in\{2,3,4\}$}}};
	\end{axis}
\end{tikzpicture}

%% file: Figures/ferRM256_37_FSC_per_rpa.tikz.tex
\begin{tikzpicture}[spy using outlines = {rectangle, magnification=3, connect spies}]

	\pgfplotsset{
		label style = {font=\fontsize{9pt}{7.2}\selectfont},
		tick label style = {font=\fontsize{7pt}{7.2}\selectfont}
	}
	
	\begin{axis}[
		scale = 1,
		ymode=log,
		xlabel={$E_b/N_0$ [\text{dB}]}, xlabel style={yshift=0.5em},
		ytick={1e-6, 1e-5,1e-4,1e-3,1e-2,1e-1,1e-0},
		xtick={1,1.5,2,2.5,3},
		ylabel={FER}, 
		grid=both,
		ymajorgrids=true,
		xmajorgrids=true,
		grid style=dashdotted,
		width=0.85\linewidth, height=6cm,
		thin,
		mark size=2.25,
		legend cell align=left,
		legend style={
			column sep= 2mm,
			font=\fontsize{7pt}{7.2}\selectfont,
		},
			legend to name=perf-legend-all-8-2,
			legend columns=4,
		]
		
		\addplot[
			color=black,
			mark=none,
			thin,
			mark size=2,
			dashdotted,
			mark options={solid},
		]
		table {
			1	0.017
			1.5	0.00504348
			2	0.00113636
			2.5	0.00017301
		};
		\addlegendentry{ML lower-bound \cite{Dumer06}}
				

		\addplot[
			color=violet,
			mark=otimes,
			thin,
			mark size=2,
			solid,
			mark options={solid},
		]
		table {
			1	3.02E-02
			1.5	9.33E-03
			2	1.87E-03
			2.5	3.41E-04
		};
		\addlegendentry{RPA \cite{Ye20}}

		\addplot[
		color=violet,
		mark=otimes,
		thin,
		mark size=2,
		dashdotted,
		mark options={solid},
		]
		table {
			1	3.60E-02
			1.5	1.13E-02
			2	2.62E-03		
			2.5	4.76E-04
		};
		\addlegendentry{SRPA \cite{fathollahi2020sparse}}
		
				\addplot[
		mark=oplus,
		thin,
		cyan,
		mark size=2,
		solid,
		mark options={solid},
		]
		table {
			1.0	2.55E-02
			1.5	8.27E-03
			2.0	1.77E-03
			2.5	2.88E-04
		};
		\addlegendentry{Aut-SSC-$P$ \cite{gabi_fast_pcd, geiselhart2020automorphism}}
		
		\addplot[
		mark=square,
		thin,
		blue,
		mark size=2,
		mark options={solid},
		]
		table {
			1	2.76E-02
			1.5	8.94E-03
			2	1.96E-03
			2.5	3.14E-04
		};
		\addlegendentry{p-FHT-FSCL-$L$}
		
		\addplot[
			mark=star,
			thin,
			red,
			mark size=2,
			mark options={solid},
		]
		table {
			1	2.63E-02
			1.5	7.76E-03
			2.0	1.84E-03
			2.5	3.01E-04
		};
		\addlegendentry{p-FHT-FSCL-$1$-$M_1$}
		
		\addplot[
			mark=triangle,
			thin,
			green,
			mark size=2,
			mark options={solid},
		]
		table {
			1	2.63E-02
			1.5	7.80E-03
			2	1.87E-03
			2.5	2.74E-04
		};
		\addlegendentry{p-FHT-FSCL-$4$-$M_4$}
					
\node[anchor=north east, fill=white, align=left] at (rel axis cs:1,1) {\footnotesize{\makecell{\normalsize{$\mathcal{RM}(2,8)$}\\$P=64, L=16$\\$M_1=25, M_4=5$}}};

\coordinate (spypoint1) at (axis cs:2, 2.2e-3);
\coordinate (magnifyglass1) at (axis cs:1.25, 0.7e-3);

\end{axis}

\spy [black, width=1.25cm, height=2cm] on (spypoint1) in node[fill=white] at (magnifyglass1);
\end{tikzpicture}

%% file: Figures/ferRM256_93_FSC_per_rpa.tikz.tex
\begin{tikzpicture}[spy using outlines = {rectangle, magnification=3, connect spies}]

	\pgfplotsset{
		label style = {font=\fontsize{9pt}{7.2}\selectfont},
		tick label style = {font=\fontsize{7pt}{7.2}\selectfont}
	}
	
	\begin{axis}[
		scale = 1,
		ymode=log,
		xlabel={$E_b/N_0$ [\text{dB}]}, xlabel style={yshift=0.5em},
		ytick={1e-6, 1e-5,1e-4,1e-3,1e-2,1e-1,1e-0},
		xtick={1,1.5,2,2.5,3,3.5},
		ylabel={FER}, 
		grid=both,
		ymajorgrids=true,
		xmajorgrids=true,
		grid style=dashdotted,
		width=0.85\linewidth, height=6cm,
		thick,
		mark size=2.25,
		legend cell align=left,
		legend style={
			column sep= 2mm,
			font=\fontsize{7pt}{7.2}\selectfont,
		},
				legend to name=perf-legend-all-8-3,
				legend columns=4,
		]

		\addplot[
			color=black,
			mark=none,
			thin,
			mark size=2,
			dashdotted,
			mark options={solid},
		]
		table {
			1	2.42000E-02
			1.5	4.04000E-03
			2	5.07614E-04
			2.5	4.80538E-05
		};
		\addlegendentry{ML (lower bound)}
		
		\addplot[
		color=violet,
		mark=otimes,
		thin,
		mark size=2,
		solid,
		mark options={solid},
		]
		table {
			1	7E-2
			1.5	1.8E-2
			2	2.4E-3
			2.5	2.2E-4
		};
		\addlegendentry{RPA \cite{Ye20}}
		
		\addplot[
			color=violet,
			mark=otimes,
			thin,
			mark size=2,
			dashdotted,
			mark options={solid},
		]
		table {
			1	7.82E-02
			1.5	2.30E-02
			2	3.57E-03
			2.5	3.88E-04
		};
		\addlegendentry{SRPA \cite{Ye20}}
		
		\addplot[
		mark=oplus,
		thin,
		cyan,
		mark size=2,
		solid,
		mark options={solid},
		]
		table {
			1	6.42E-02
			1.5	1.17E-02
			2	1.62E-03
			2.5	1.25E-04
		};
		\addlegendentry{Aut-FSC-$P$ \cite{geiselhart2020automorphism}}
		
		
		
		\addplot[
		mark=star,
		thin,
		red,
		mark size=2,
		mark options={solid},
		]
		table {
			1	7.54E-02
			1.5	1.63E-02
			2	2.29E-03
			2.5	1.53E-04
		};
		\addlegendentry{p-FHT-FSCL-$1$-$M_1$}
		
		\addplot[
		mark=triangle,
		thin,
		green,
		mark size=2,
		mark options={solid},
		]
		table {
			1.0	6.55E-02
			1.5	1.42E-02
			2.0	1.69E-03
			2.5	1.15E-04
		};
		\addlegendentry{p-FHT-FSCL-$4$-$M_4$}
		
		\addplot[
		mark=square,
		thin,
		blue,
		mark size=2,
		mark options={solid},
		]
		table {
			1	5.45E-02
			1.5	1.11E-02
			2.0	1.42E-03
			2.5	1.28E-04
		};
		\addlegendentry{p-FHT-FSCL-$L$}
		
		\node[anchor=north east, fill=white, align=right] at (rel axis cs:1,1) {\footnotesize{\makecell{\normalsize{$\mathcal{RM}(3,8)$}\\$P=256, L=64$\\$M_1=100, M_4=16$}}};

\coordinate (spypoint1) at (axis cs:2.0, 2.2e-3);
\coordinate (magnifyglass1) at (axis cs:1.25, 3e-4);

\end{axis}

\spy [black, width=1.25cm, height=2cm] on (spypoint1) in node[fill=white] at (magnifyglass1);
\end{tikzpicture}

%% file: Figures/ferRM256_163_FSC_per_rpa.tikz.tex
\begin{tikzpicture}[spy using outlines = {rectangle, magnification=3, connect spies}]

	\pgfplotsset{
		label style = {font=\fontsize{9pt}{7.2}\selectfont},
		tick label style = {font=\fontsize{7pt}{7.2}\selectfont}
	}
	
	\begin{axis}[
		scale = 1,
		ymode=log,
		xlabel={$E_b/N_0$ [\text{dB}]}, xlabel style={yshift=0.5em},
		ytick={1e-6, 1e-5,1e-4,1e-3,1e-2,1e-1,1e-0},
		xtick={2,2.5,3,3.5},
		ylabel={FER}, 
		grid=both,
		ymajorgrids=true,
		xmajorgrids=true,
		grid style=dashdotted,
		width=0.85\linewidth, height=6cm,
		thick,
		mark size=2.25,
		legend cell align=left,
		legend style={
			column sep= 2mm,
			font=\fontsize{7pt}{7.2}\selectfont,
		},
				legend to name=perf-legend-4-8,
				legend columns=4,
		]
		
		\addplot[
			color=black,
			mark=none,
			thin,
			mark size=2,
			dashdotted,
			mark options={solid},
		]
		table {
			2	2.21000E-02
			2.5	2.71053E-03
			3	2.31690E-04
			3.5	1.51926E-05
		};
		
		\addplot[
			color=violet,
			mark=otimes,
			thin,
			mark size=2,
			solid,
			mark options={solid},
		]
		table {
			2	8.53E-2 
			2.5	1.58E-2 
			3	1.35E-3 
			3.5	8.00E-5
		};
		\addlegendentry{RPA \cite{Ye20}}
		
				\addplot[
		color=violet,
		mark=otimes,
		thin,
		mark size=2,
		dashdotted,
		mark options={solid},
		]
		table {
			2	1.36E-1
			2.5	2.80E-2
			3	3.34E-3
			3.5	2.26E-04
		};
		
		\addplot[
		mark=oplus,
		thin,
		cyan,
		mark size=2,
		solid,
		mark options={solid},
		]
		table {
			2	8.43E-02
			2.5	1.29E-02
			3	1.01E-03
			3.5 5.15E-05
		};
		

		\addplot[
			mark=star,
			thin,
			red,
			mark size=2,
			mark options={solid},
		]
		table {
			2	7.63E-02
			2.5	1.14E-02
			3.0	8.28E-04
			3.5	4.08E-05
		};
		\addlegendentry{p-FHT-FSCL-$1$-${M}$}
		
		\addplot[
			mark=triangle,
			thin,
			green,
			mark size=2,
			mark options={solid},
		]
		table {
			2	6.70E-02
			2.5	9.78E-03
			3	7.36E-04
			3.5	3.37E-05
		};
		\addlegendentry{p-FHT-FSCL-$4$-$16$}
		
		\addplot[
		mark=square,
		thin,
		blue,
		mark size=2,
		mark options={solid},
		]
		table {
			2.0	5.89E-2
			2.5	8.07E-03
			3.0	5.34E-04
			3.5	2.44E-05
		};
		\addlegendentry{p-FHT-FSCL-$L$}
		
		\node[anchor=north east, fill=white, align=right] at (rel axis cs:1,1) {\footnotesize{\makecell{\normalsize{$\mathcal{RM}(4,8)$}\\$P=128, L=64$\\$M_1=80, M_4=16$}}};

\coordinate (spypoint1) at (axis cs:3.0, 0.8e-3);
\coordinate (magnifyglass1) at (axis cs:2.25, 2e-4);

\end{axis}

\spy [black, width=1.25cm, height=2cm] on (spypoint1) in node[fill=white] at (magnifyglass1);
\end{tikzpicture}

%% file: Figures/ferRM512_46_SSC_per_rpa.tikz.tex
\begin{tikzpicture}[spy using outlines = {rectangle, magnification=3, connect spies}]

	\pgfplotsset{
		label style = {font=\fontsize{9pt}{7.2}\selectfont},
		tick label style = {font=\fontsize{7pt}{7.2}\selectfont}
	}
	
	\begin{axis}[
		scale = 1,
		ymode=log,
		xlabel={$E_b/N_0$ [\text{dB}]}, xlabel style={yshift=0.5em},
		ytick={1e-7, 1e-6, 1e-5,1e-4,1e-3,1e-2,1e-1,1e-0},
		xtick={1,1.5,2,2.5,3},
		ylabel={FER}, 
		grid=both,
		ymajorgrids=true,
		xmajorgrids=true,
		grid style=dashdotted,
		width=0.85\linewidth, height=6cm,
		thin,
		mark size=2.25,
		legend cell align=left,
		legend style={
			column sep= 2mm,
			font=\fontsize{7pt}{7.2}\selectfont,
		},
		legend to name=perf-legend-FSC-FHT-9-2,
		legend columns=4,
		]
		
		\addplot[
		color=violet,
		mark=otimes,
		thin,
		mark size=2,
		dashdotted,
		mark options={solid},
		]
		table {
			1.0	1.73E-02
			1.5	4.06E-03
			2.0	5.00E-04
			2.5	5.10E-05
		};
		
		\addplot[
		color=black,
		mark=none,
		thin,
		mark size=2,
		dashdotted,
		mark options={solid},
		]
		table {
			1	5.37E-3
			1.5	8.87E-4
			2	1.5E-4
			2.5	2.01E-5
		};
		

		\addplot[
			color=violet,
			mark=otimes,
			thin,
			mark size=2,
			solid,
			mark options={solid},
		]
		table {
			1	1.18E-02
			1.5	2.09E-03
			2	2.81E-04
			2.5	2.53E-05	
		};
		\addlegendentry{RPA \cite{Ye20}}
		
		\addplot[
			mark=oplus,
			thin,
			cyan,
			mark size=2,
			solid,
			mark options={solid},
		]
		table {
			1.0	9.38E-03
			1.5	2.14E-03
			2.0	2.58E-04
			2.5	2.76E-05
		};
		\addlegendentry{Aut-FSC-$512$ \cite{geiselhart2020automorphism}}
		
		\addplot[
		mark=star,
		thin,
		red,
		mark size=2,
		mark options={solid},
		]
		table {
			1	1.04E-02
			1.5	2.34E-03
			2.0	2.74E-04
			2.5	2.54E-05
		};
		\addlegendentry{p-FHT-FSCL-$1$-$100$}
		
		\addplot[
			mark=triangle,
			thin,
			green,
			mark size=2,
			mark options={solid},
		]
		table {
			1.0	9.80E-03
			1.5	2.11E-03
			2.0	2.99E-04
			2.5	2.85E-05
		};
		\addlegendentry{p-FHT-FSCL-$4$-$\frac{M}{4}$}
		
		\addplot[
			mark=square,
			thin,
			blue,
			mark size=2,
			mark options={solid},
		]
		table {
			1	9.55E-03
			1.5	2.11E-03
			2.0	2.78E-04
			2.5 2.87E-05
		};
		\addlegendentry{p-FHT-FSCL-$64$}
		
		\node[anchor=north east, fill=white, align=right] at (rel axis cs:1,1) {\footnotesize{\makecell{\normalsize{$\mathcal{RM}(2,9)$}\\$P=512, L=64$\\$M_1=100, M_4=20$}}};

\coordinate (spypoint1) at (axis cs:2, 3.4e-4);
\coordinate (magnifyglass1) at (axis cs:1.25, 1e-4);

\end{axis}

\spy [black, width=1.25cm, height=2cm] on (spypoint1) in node[fill=white] at (magnifyglass1);
\end{tikzpicture}